\documentclass[fleqn,usenatbib]{mnras}

\usepackage[T1]{fontenc}
\DeclareRobustCommand{\VAN}[3]{#2}
\let\VANthebibliography\thebibliography
\def\thebibliography{\DeclareRobustCommand{\VAN}[3]{##3}\VANthebibliography}

\usepackage{graphicx}	
\usepackage{amsmath}	
\usepackage{amssymb}	
\usepackage{appendix}
\usepackage{newtxtext,newtxmath}

\title[Hybrid WD mergers]{Thermonuclear explosion of a massive hybrid HeCO white-dwarf triggered by a He-detonation on a companion}

\author[R. Pakmor et al.]{
R. Pakmor,$^{1}$\thanks{E-mail: rpakmor@mpa-garching.mpg.de}
Y. Zenati$^{2}$,
H. B. Perets$^{2}$,
S. Toonen$^{3,4}$
\\
$^1$Max-Planck-Institut f\"{u}r Astrophysik, Karl-Schwarzschild-Str. 1, D-85748, Garching, Germany\\
$^{2}$Physics Department, Technion - Israel Institute of Technology, Haifa 3200004, Israel\\
$^{3}$Institute of Gravitational Wave Astronomy, School of Physics and Astronomy, University of Birmingham, Birmingham, B15 2TT,United Kingdom\\
$^{4}$Anton Pannekoek Institute for Astronomy, University of Amsterdam, 1090 GE Amsterdam, The Netherlands\\
}

\date{Accepted XXX. Received YYY; in original form ZZZ}

\pubyear{2020}

\begin{document}
\label{firstpage}
\pagerange{\pageref{firstpage}--\pageref{lastpage}}
\maketitle

\begin{abstract}
Normal type Ia supernovae (SNe) are thought to arise from the thermonuclear explosion of massive ($>0.8$ M$_\odot$) carbon-oxygen white dwarfs (WDs), although the exact mechanism is debated. In some models helium accretion onto a carbon-oxygen (CO) WD from a companion was suggested to dynamically trigger a detonation of the accreted helium shell. The helium detonation then produces a shock that after converging on itself close to the core of the CO-WD, triggers a secondary carbon detonation and gives rise to an energetic explosion. However, most studies of such scenarios have been done in one or two dimensions, and/or did not consider self-consistent models for the accretion and the He-donor. Here we make use of detailed 3D simulation to study the interaction of a He-rich hybrid $0.69\,\mathrm{M_\odot}$ HeCO WD with a more massive $0.8\,\mathrm{M_\odot}$ CO~WD. We find that accretion from the hybrid WD onto the CO~WD gives rise to a helium detonation. However, the helium detonation does not trigger a carbon detonation in the CO~WD. Instead, the helium detonation burns through the accretion stream to also burn the helium shell of the donor hybrid HeCO-WD. The detonation of its massive helium shell then compresses its CO core, and triggers its detonation and full destruction. The explosion gives rise to a faint, likely highly reddened transient, potentially observable by the Vera Rubin survey, and the high-velocity ($\sim 1000\,\mathrm{km s^{-1}}$) ejection of the heated surviving CO~WD companion. Pending on uncertainties in stellar evolution we estimate the rate of such transient to be up to $\sim10\%$ of the rate of type Ia SNe.
\end{abstract}

\begin{keywords}
stars: binaries -- supernovae: general -- hydrodynamics -- nucleosynthesis -- transients: supernovae
\end{keywords}



\section{Introduction}
It was recently suggested that hybrid HeCO~WDs might play a key role in the production of thermonuclear supernovae (SNe)\citep{PeretsZenati2019,Zenati2019}. Their disruption by a more massive CO~WD, and later accretion of the debris onto the CO~WD was shown in 2D simulations to trigger a thermonuclear explosion of the CO~WD, giving rise to a type Ia supernova. 
However, due to the limitations of 2D simulations, these models did not simulate the early phases of accretion from the HeCO~WD onto the CO~WD and the later disruption. In particular, accretion of helium from the hybrid WD onto the CO~WD could potentially give rise to a dynamical detonation of the accreted helium layer (or a pre-existing helium layer on the CO~WD if such exists). In such a case it is possible that the helium layer detonation may trigger a secondary carbon detonation of the CO~WD leading to its explosion, even before the companion hybrid WD is disrupted. This latter possibility follows ideas of double-detonation models beginning in the 80's \cite{Ibe85,Iben+87}, and their more recent incarnation as dynamical double detonations \cite[see, e.g.]{Guillochon2010,Pakmor2013,Sato+15,Shen2018b}.

In order to understand the outcomes of a CO-HeCO double degenerate binary system, and explore whether it gives rise to the disruption of the HeCO~WD, we employ dynamical 3D simulations that begins following the final phase of the inspiral of the binary CO-HeCO~WD system.

We specifically investigate the fate of a double degenerate binary system consisting of a primary pure CO-WD with a mass of $0.8\,\mathrm{M_\odot}$ and a secondary hybrid HeCO WD with total mass of $0.69\,\mathrm{M_\odot}$ made of a CO core of $0.59\,\mathrm{M_\odot}$ and a massive helium shell of $0.1\,\mathrm{M_\odot}$. Note that both the mass ratio, and the highly He-enriched HeCO WD {\it differ} from those studied in 2D accretion disk simulations \cite{PeretsZenati2019}. In particular, the much higher He shell for this hybrid is attained through somewhat different evolution than that described in \cite{Zenati2019}, as we further discuss in section 2 below.

We obtain the density profile and composition of both WDs using the 1D stellar evolution code \textsc{mesa} \citep{Paxton2011,Paxton2019} similarly to \citet{Zenati2019}. We follow the evolution of the hybrid WD starting from the main sequence as a $4.1\mathrm{M_\odot}$ star in a binary system with a separation of $a=4.18\,\mathrm{AU}$ with a more massive $7.5\,\mathrm{M_\odot}$ companion star that will become the primary WD, and adopt solar metallicity $\rm Z=Z_{\odot}=0.02$. We discuss the formation of the double degenerate binary system in detail in Sec.~\ref{sec:binary}~and~\ref{sec:popsynth}.

Starting from the 1D profiles we then generate 3D representations of these WDs in the moving mesh hydrodynamics code \textsc{arepo} \citep{Arepo,Pakmor2016} that includes a fully coupled nuclear reaction network. We initialise the binary system with an initial separation of $4\times 10^4\,\mathrm{km}$. We follow the inspiral of the binary system as the secondary hybrid WD fills its Roche lobe and transfers material to the primary WD. Eventually a detonation ignites on the surface of the primary WD in its accreted helium shell. We describe the inspiral phase and ignition of the helium detonation in Sec.~\ref{sec:inspiral}.

The helium detonation sweeps around the primary CO~WD and sends a shock wave into its center. As there is not a lot of helium around the primary WD the detonation is weak. Its shock wave still converges at the edge of the CO core of the primary WD but fails to detonate carbon. In contrast, at the same time the helium detonation propagates upwards through the dense accretion stream towards the secondary hybrid WD. When it arrives there the helium detonation also travels around the hybrid WD, burning its massive helium shell. Here the detonation is strong and fast and its shock wave converges close to the center of the hybrid WD where it manages to ignite the CO core. The emerging carbon detonation then burns and unbinds the whole secondary WD. We describe this phase and characterise the ejecta of the secondary hybrid WD and the properties of the surviving primary CO~WD in Sec.~\ref{sec:explosion}.

We then employ the radiation transfer code SuperNu \citep{Wollaeger2013,Wollaeger2014} to compute synthetic lightcurves for the explosion in Sec.~\ref{sec:observables}. We use population synthesis to estimate the frequency of similar events and discuss the observability of the explosion and the surviving WDs in Sec.~\ref{sec:popsynth}.

We finally summarise our results and provide an outlook on the most important questions that arise from our work in Sec.~\ref{sec:conclusion}.

\section{Formation of the binary system}
\label{sec:binary}

At the current age of the Universe single star evolution has produced CO~WDs in the mass range $\sim 0.50$ - $\rm 1.05\ M_\odot$ and oxygen-neon (ONe) WDs in the mass range $\sim 1.05$ - $\rm 1.38\ M_\odot$. However, binary evolution makes this picture much more complex and can give rise to WDs with very different properties including very low mass (VLM) WDs \citep{Ist+16,Zha+18}. Moreover, there are two ways how binary systems can produce hybrid WDs with a CO core and an outer helium shell, either through a phase of mass transfer via Roche-lobe overflow (RLOF) or through a common envelope phase (CEE) \cite[see e.g.][]{Iva13}.

During this binary interaction the hydrogen-rich envelope of the star that will become the hybrid WD is stripped following the formation of a He core. The later evolution of the stripped star and its helium core is then significantly altered compared to the uninterrupted evolution of an identical but non-interacting (single) star. After most of the red giant envelope is removed the outer hydrogen shell burning is quelled, but the helium-core keeps growing in mass \citep{Moroni09}, and the star begins to contract. If the helium core is sufficiently massive, the contraction will eventually trigger helium ignition \citep{Ibe85,Zenati2019} and the formation of a CO core. In this case the helium to CO abundance ratio of the final WD will be determined by the specific detailed evolution of helium burning as well as mass-loss through winds from the envelope \citep{Tut+92,Moroni09,Zenati2019}.

As shown in \citet{Zenati2019} such interacting binary systems can produce hybrid WDs in the mass range of $\rm 0.38-$  $\rm 0.72 M_\odot$ with a helium envelope containing $\rm \sim 2-$ $20\%$ of the total mass of the WD. Recent observational evidence for the existence of such hybrid WDs has been mounting. The ZTF survey \citep{Kupfer20} reported their finding of the first short-period binary in which a hot subdwarf star (sdOB) has filled its Roche lobe and has started mass transfer to its companion. The binary system has a period of $P = 39.3401\,\mathrm{min}$, making it the most compact hot subdwarf binary currently known. \citet{Kupfer20} estimated that the hot subdwarf will become a hybrid WD (with a helium layer of $\sim 0.1\,\mathrm{M_\odot}$) and merge with its CO~WD companion in about $17$~Myr. In this case it may end in a thermonuclear explosion or form an R~CrB star. In addition \cite{Beuermann20} and \cite{Steven20} found eclipsing binaries that may have a hybrid WD as their primary star.

\subsection{A highly He-enriched hybrid WD}
In this work we explore the fate of a close binary system consisting of a $0.8\,\mathrm{M_\odot}$ CO~WD and a $0.69\,\mathrm{M_\odot}$ hybrid WD that has a helium shell with a mass of $0.1\,\mathrm{M_\odot}$ or $\sim 14\%$ of its mass. 
Although this hybrid WD is produced through the same stellar evolution stages as the hybrid WD of the same total mass in \cite{Zenati2019}, some different choices of parameters give rise to a more He-enriched WD than those described in \cite{Zenati2019}. We follow the stellar evolutionary tracks of both binary components from the pre-main sequence stage to the final binary system consisting of two WDs. We stop the evolution once the star becomes a fully degenerate WD. This condition effectively translates to a WD luminosity and temperature below $L \leq 1.12\ L_\mathrm{\odot}$ and $T_\mathrm{eff} \leq 4.92 \ T_\mathrm{\odot,eff}$, respectively.

The evolution of the binary system depends strongly on the initial conditions. Based on our population modelling, we describe the typical binary evolution in Section\,\ref{sec:popsynth}. It is important to note that the initial conditions of the binary system studied here are {\rm different} than those described in \cite{Zenati2019} for the formation of the $0.69$ M$_\odot$ HeCO~WD. Here we begin with an initial mass ratio $\rm q= M_{donor}/M_{companion} \sim 0.58$ and an initial orbital period $\rm P=4.37\,\mathrm{d}$. We also consider slightly different overshooting parameter (0.0012) and mixing parameter ($\alpha=1.3$)

We explored a range of hybrid WDs to chose the parameters and found these parameters give rise to very similar results for WDs less massive than $0.63$ M$_\odot$ as described in \citep{Zenati2019}. However, when considering these parameters for the formation of more massive hybrids, they give rise to the formation of an even more He-enriched HeCO~WD than discussed in \citet{Zenati2019}, such as the one considered here. These results will be discussed in depth in a dedicated paper. 

\begin{figure}
\includegraphics[width=0.97\linewidth]{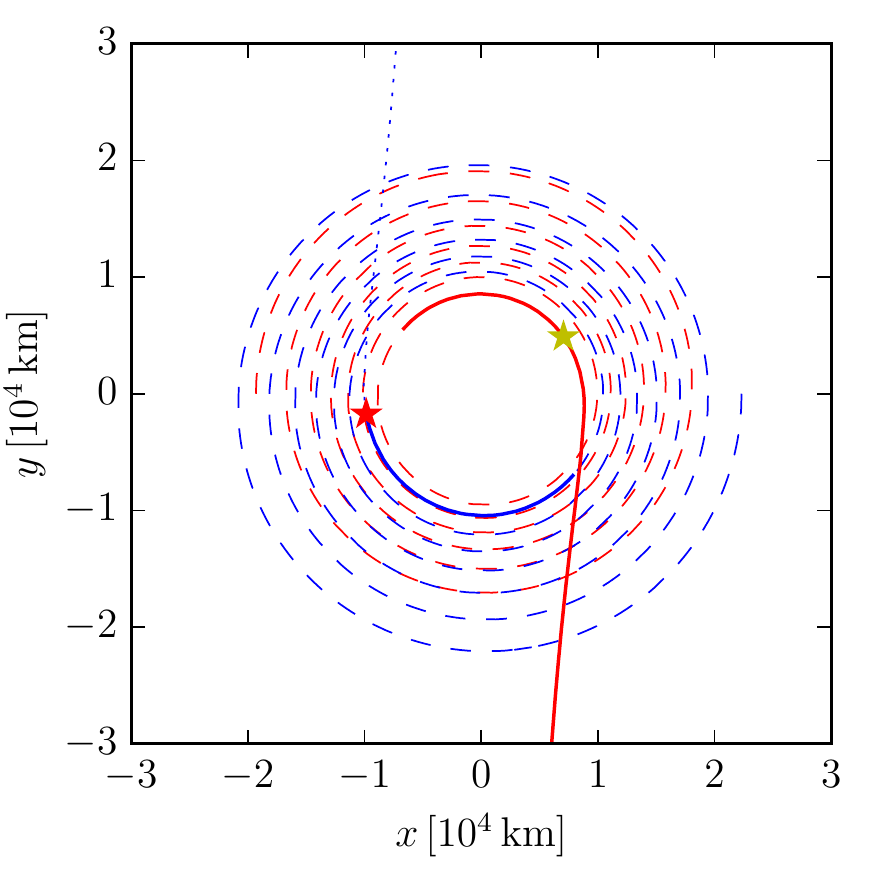}
\includegraphics[width=0.97\linewidth]{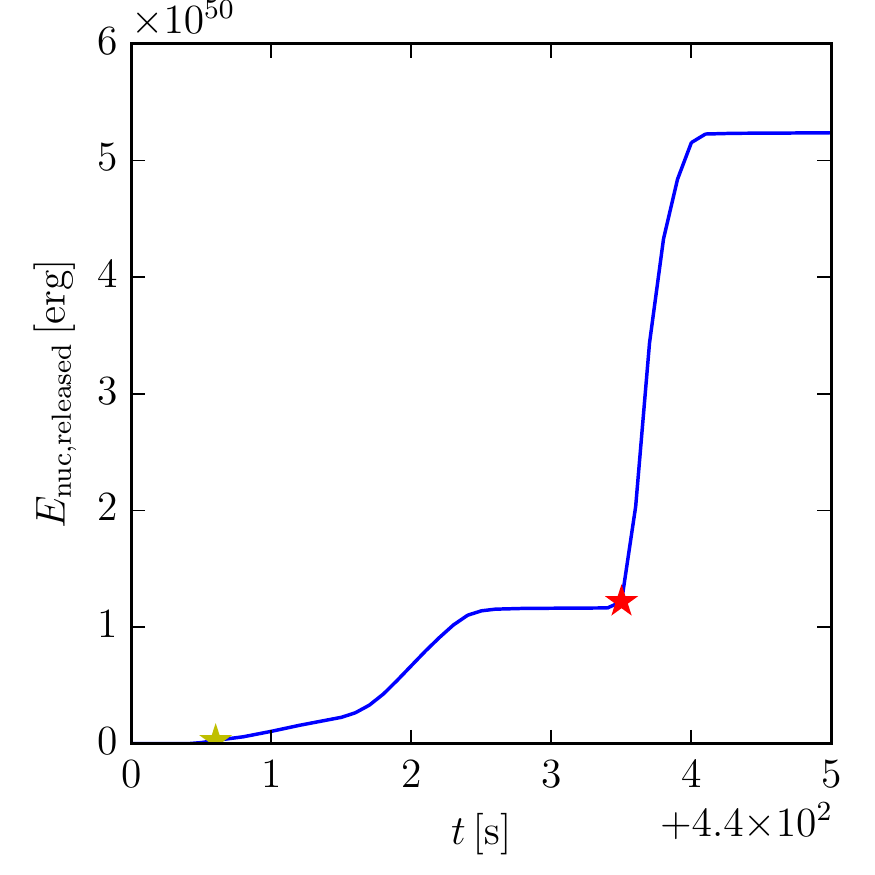}

\caption{The top panel shows the orbital evolution of the two WDs in the binary system in the plane of rotation. Dashed lines show their path while the additional force mimicking gravitational wave emission is active. Straight lines show the evolution of the primary WD (red) and the secondary WD (blue). The latter ends when the secondary WD is disrupted. The yellow star denotes the moment when the helium detonation ignites first on the surface of the primary WD. The red star marks the time when the carbon detonation ignites in the center of the secondary WD after which it is quickly disrupted. The blue dotted line shows the movement of the center of mass of the ejecta of the secondary WD. The lower panel shows the cumulative total nuclear energy released. Stars again mark the ignition of the helium detonation (yellow) and the carbon detonation (red).}
\label{fig:evolution}
\end{figure}

\begin{figure*}
\includegraphics[width=0.81\textwidth]{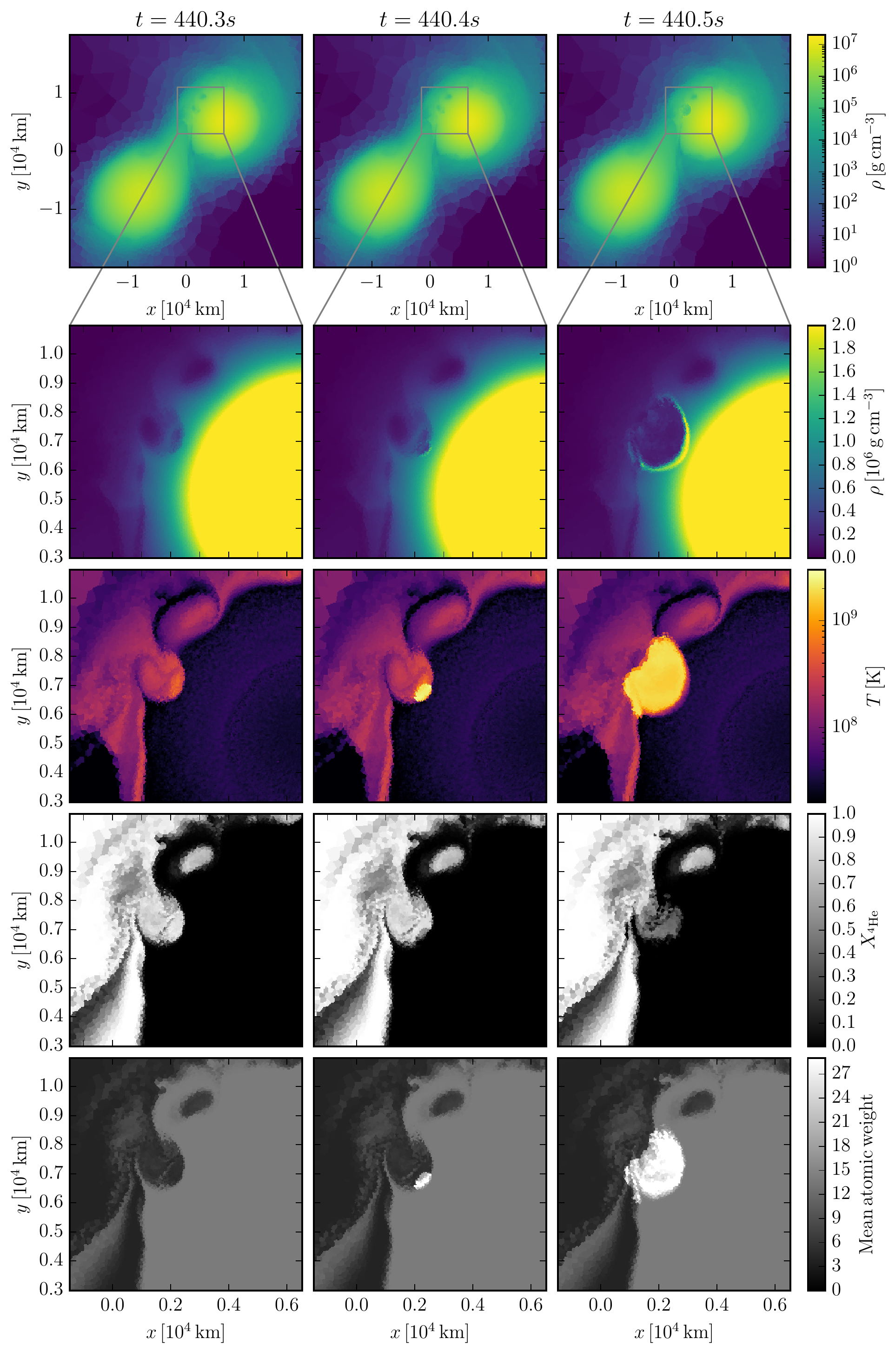}
\caption{Columns show slices through the midplane of the binary for the snapshots just before, at, and directly after the formation of the first detonation, the helium detonation on the surface of the primary CO~WD. The snapshots are separated by $0.1\mathrm{s}$. The top row shows a density slice of the whole binary, the bottom rows show slices zooming in on the position where the detonation forms featuring density, temperature, mass fraction of $^4\mathrm{He}$, and mean atomic weight, respectively.}
\label{fig:he_det}
\end{figure*}

\section{Inspiral and ignition}
\label{sec:inspiral}

Once a binary system with two WD components has formed, it slowly loses angular momentum via gravitational wave emission until eventually the two WDs get sufficiently close for the secondary (hybrid) WD to fill its Roche lobe. It then starts transferring mass onto the primary CO~WD. For a long time mass transfer is very slow and irrelevant for the properties of the system, while the orbit continues to shrink from further emission of gravitational waves, and the mass transfer rate increases. Eventually the binary becomes sufficiently close such that the mass transfer rate is large enough as to transfer substantial amounts of mass on a timescale of several orbits. 
At this point, which serves as the initial starting point of our 3D simulation, the density of the accretion stream continues to increase, and unless interrupted by other process would eventually lead to the disruption of the secondary, after a few tens of orbits. However, we find that in the configuration explored here, thermonuclear detonation occurs beforehand, and gives rise to very different outcomes, as we now describe.

\begin{figure*}
\includegraphics[width=\textwidth]{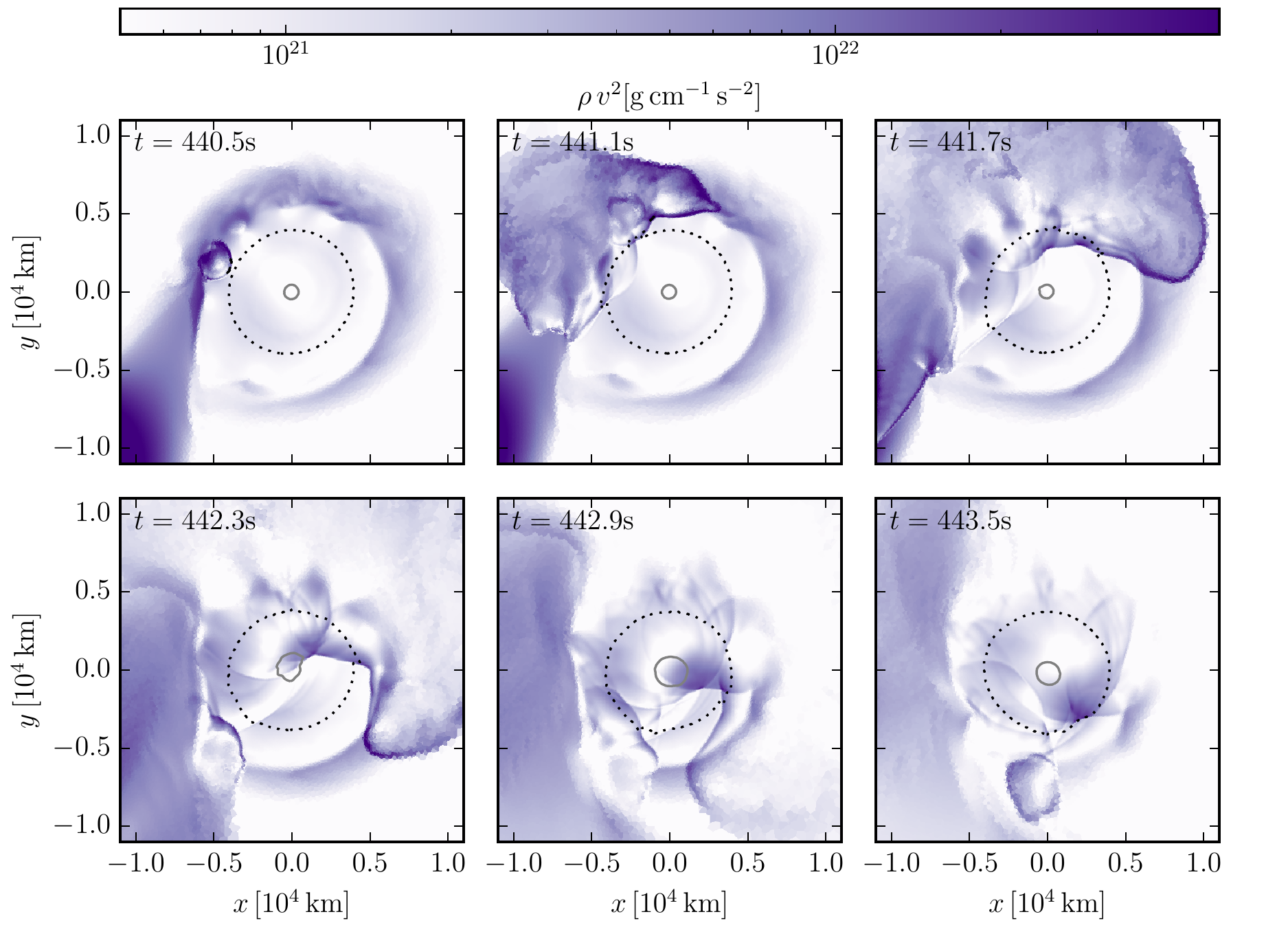}
\caption{Propagation of the shock originating from the helium detonation as it moves around the primary WD. The panels show the time evolution from the time of the ignition of the helium detonation (top left panel) to the time when the shock converges in the CO core of the primary WD (bottom right panel). The black dotted and gray solid contours indicate densities of $2\times 10^6\,\mathrm{g\,cm^{-3}}$ and $10^7\,\mathrm{g\,cm^{-3}}$, respectively.}
\label{fig:shock_pri}
\end{figure*}

\begin{figure*}
\includegraphics[width=\textwidth]{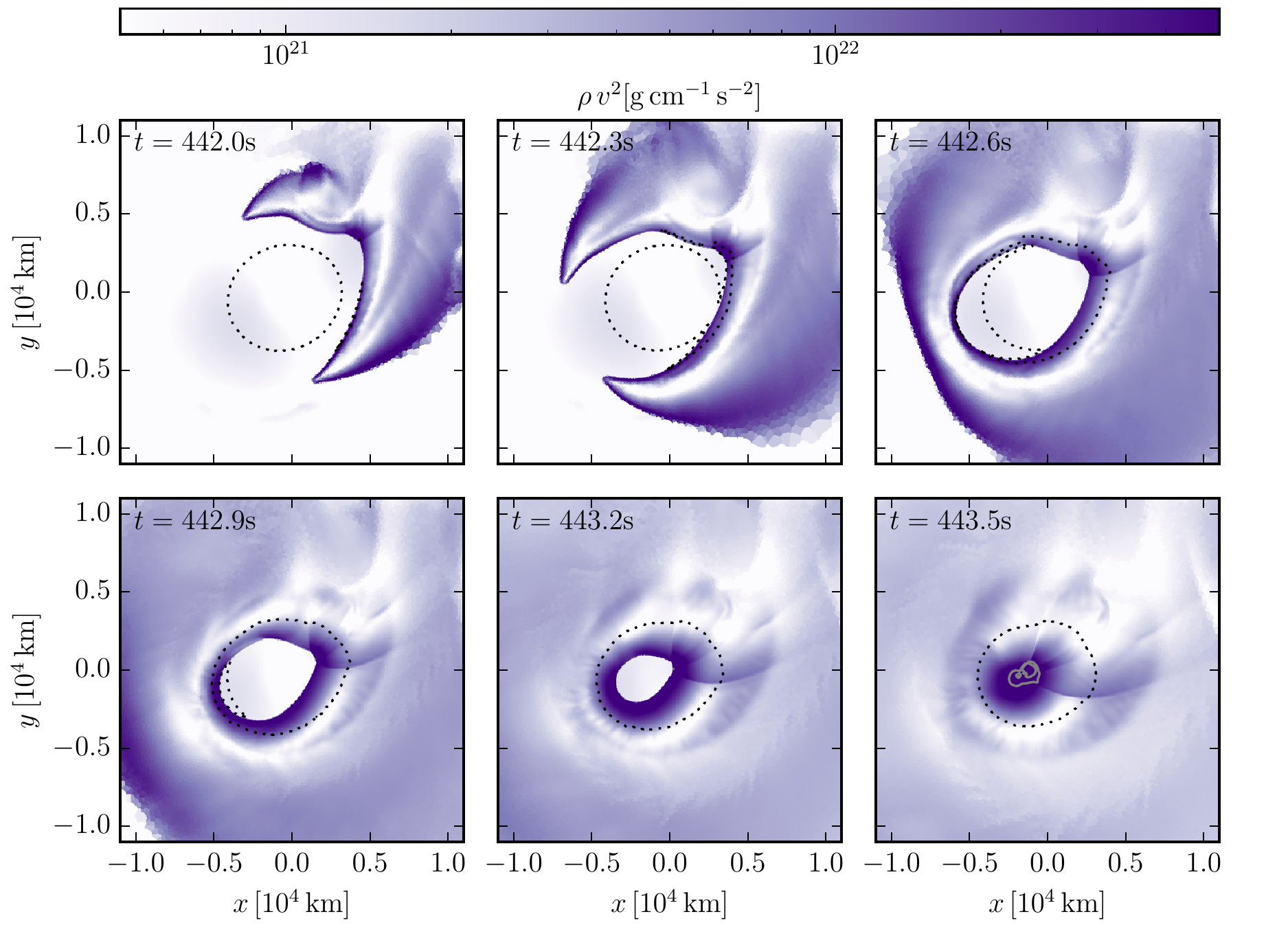}
\caption{Propagation of the shock originating from the helium detonation moves around the seconday WD. The panels show the time evolution from the time when the helium detonation arrives at the secondary WD and starts to propagate around it (top left panel) to the time when the shock converges in the carbon-oxygen core of the secondary WD (bottom right panel).  The black dotted and gray solid contours indicate densities of $2\times 10^6\,\mathrm{g\,cm^{-3}}$ and $10^7\,\mathrm{g\,cm^{-3}}$, respectively.}
\label{fig:shock_sec}
\end{figure*}

\subsection{Arepo}

To model the dynamical evolution of the binary system we use the moving-mesh code \textsc{arepo} \citep{Arepo,Pakmor2016,Weinberger2020}. It discretises the volume into cells and solves the equations of hydrodynamics using a second order finite volume scheme \citep{Pakmor2016}. Its Voronoi mesh is reconstructed in every timestep from a set of mesh-generating points that each span up single cell. Fluxes over interfaces are computed using the HLLC Riemann solver in the moving frame of the interface \citep{Pakmor2011b}. We use \textsc{arepo} in its pseudo-Lagrangian mode, i.e.the mesh-generating points follow the gas velocity with small corrections to keep the mesh regular. On top of the movement of the mesh-generating points we employ explicit refinement and de-refinement for cells that are more than a factor of two away from the desired target gas mass of the cells. For the simulation presented here, our target mass resolution is always $m_\mathrm{target}=10^{-7}\,\mathrm{M_\odot}$. In addition we require that the volume of a cell is not more than $10$ times larger than its largest direct neighbour, otherwise the cell is refined.

In addition to hydrodynamics \textsc{arepo} includes self-gravity of the gas. Gravitational accelerations are computed using a tree solver and are coupled to the hydrodynamics via a Leapfrog time integration scheme. The gravitational softening of the cells is set to be equal to $2.8$ times their radius, with a minimum softening of $10\,\mathrm{km}$. To improve the efficiency of the simulation we use local timesteps in \textsc{arepo}, i.e. every cell is integrated on the largest timestep of a discrete set of timesteps, that is smaller than the timestep of its local timestep criteria. This way only a small number of cells is integrated on the smallest timestep in the simulation and the bulk of the cells can be integrated on much larger timesteps.

To model degenerate electron gases present in WDs we use the \textsc{helmholtz} equation of state \citep{Timmes2000} including Coulomb corrections. Moreover, we include a $55$~isotope nuclear reaction network fully coupled to the hydrodynamics \citep{Pakmor2012}. The included isotopes are $\mathrm{n}$, $\mathrm{p}$, $^4\mathrm{He}$, $^{11}\mathrm{B}$, $^{12-13}\mathrm{C}$, $^{13-15}\mathrm{N}$, $^{15-17}\mathrm{O}$, $^{18}\mathrm{F}$, $^{19-22}\mathrm{Ne}$, $^{22-23}\mathrm{Na}$, $^{23-26}\mathrm{Mg}$, $^{25-27}\mathrm{Al}$, $^{28-30}\mathrm{Si}$, $^{29-31}\mathrm{P}$, $^{31-33}\mathrm{S}$, $^{33-35}\mathrm{Cl}$, $^{36-39}\mathrm{Ar}$, $^{39}\mathrm{K}$, $^{40}\mathrm{Ca}$, $^{43}\mathrm{Sc}$, $^{44}\mathrm{Ti}$, $^{47}\mathrm{V}$, $^{48}\mathrm{Cr}$, $^{51}\mathrm{Mn}$, $^{52,56}\mathrm{Fe}$, $^{55}\mathrm{Co}$, $^{56,58-59}\mathrm{Ni}$. We use the JINA reaction rates \citep{Cyburt2010}. Nuclear reactions are computed for all cells with $T > 10^6\,\mathrm{K}$ except for cells that are part of the shock front which we assume to be the case when $\nabla \cdot \vec{v} < 0$ and $\left| \nabla P \right| r_\mathrm{cell} / P > 0.66$ \citep{Seitenzahl2009}. Note that we reran the simulation until the nuclear burning ceases with an additional limiter that artificially reduces nuclear reaction rates to guarantee that the nuclear timescale is always longer than the hydrodynamical timestep of a cell \citep{Kushnir2013,Shen2018}. As we show and discuss in appendix~\ref{app} the results are essentially identical, with the main difference that the helium detonation moves slower with the additional limiter that reduces the reaction rates.

\subsection{Setup and Inspiral}

From the stellar evolution calculation we obtain the density profile, temperature profile, and composition profile of both WDs. To generate the 3D initial conditions in \textsc{arepo} we employ a healpix based algorithm that generates roughly cubical initial cells \citep{Pakmor2012,Ohlmann2017}. We first put both stars individually into a box with boxsize $10^5\,\mathrm{km}$ with a background density of $10^{-5}\,\mathrm{g\,cm^{-3}}$ and background specific thermal energy of $10^{14}\,\mathrm{erg\,g^{-1}}$. We relax them for $40\,\mathrm{s}$ (for the $0.8\,\mathrm{M_\odot}$ CO~WD) and $60\,\mathrm{s}$ (for the $0.69\,\mathrm{M_\odot}$ hybrid WD. For the first $80\%$ of the relaxation we apply a friction force that damps out initial velocities that are introduced from noise in the original mesh. In the last $20\%$ of the relaxation time we disable the friction force and check that the density profiles of the WDs do not change anymore, i.e. that the relaxed stars are stable at their initial profile.

After the relaxation we take the final state of both WDs from their relaxation in isolation and add them together into a simulation box with a size of $10^7\,\mathrm{km}$. We put the WDs on a spherical co-rotating orbit at a distance of $a=4.2\times10^4\,\mathrm{km}$, which sets the initial orbital period to $T=120\,\mathrm{s}$. We chose this initial orbit rather wide so that the initial tidal forces of the WDs on each other are small. We use a passive tracer fluid to track the material of each WD individually in the simulation. This also allows us to easily compute the centers of both WDs.

At this separation gravitational wave inspiral is still relevant, though it takes too much time to follow the system with its true inspiral rate. To circumvent this problem we add a tidal force that removes angular momentum from the binary system similar to gravitational waves. However, we chose the force such that the separation $a$ decreases at a constant rate $v_\mathrm{a}$, i.e.
\begin{equation}
\frac{da}{dt} = v_\mathrm{a}.
\end{equation}

This leads to a purely azimuthal acceleration on the primary WD given by
\begin{equation}
\vec{a}_{1} = - \frac{M_{2}^2}{M_1+M_2} \frac{G}{2a} \frac{\vec{v}_{1}}{\vec{v}_{1}^2}  v_\mathrm{a},
\end{equation}
and vice versa for the secondary WD. We chose $v_\mathrm{a}=50\,\mathrm{km\,s^{-1}}$ so that the orbit changes fast compared to physical gravitational wave emission but slowly compared to the dynamical timescales of the WDs so that the WDs can easily adapt to the changing tidal forces.

The evolution of the orbits of the two WDs is shown in Fig.~\ref{fig:evolution}. At $t=430\,\mathrm{s}$, after about six orbits, the orbital period has decreased to $T=38\,\mathrm{s}$ and the separation to $a=1.9\times10^4\,\mathrm{km}$. At this point we switch off the angular momentum loss term. The point where it is switched off is chosen such that the accretion stream is dense enough to dynamically affect the surface of the primary WD. 

At this time the secondary WD has donated $4\times10^{-3}\,\mathrm{M_\odot}$ of helium to the primary WD which now forms a helium shell on the surface of the primary WD. Moreover, about $4\times 10^{-5}\mathrm{M_\odot}$ of pure helium material has been unbound from the system through tidal tails.

\subsection{Ignition of the helium detonation}

A density slice through the midplane of the binary system is shown in top row of Fig.~\ref{fig:he_det} when the system has evolved $10\,\mathrm{s}$ past the point when we switched off the angular momentum loss term. As the accretion stream shears along the surface of the primary CO~WD it generates Kelvin-Helmholtz instabilities that disturb the initially separated helium and carbon-oxygen layers. At this time the accreted helium layer and the CO core of the primary WD have mixed only very little.


A zoom-in on this interface (additional rows in Fig.~\ref{fig:he_det}) shows the base of one helium rich bubble on the surface of the primary CO~WD with a radius of about $10^3\,\mathrm{km}$.
The gas here is compressed and heated up and eventually reaches a temperature of about $10^9\,\mathrm{K}$ at a density larger than $2\times10^{5}\,\mathrm{g\,cm^{-3}}$. The cells in this hotspot have a typical radius of $15\,\mathrm{km}$. Under these conditions helium starts to burn explosively and a helium detonation forms quickly, consistent with ignition simulations of resolved helium detonations \citep{Shen2014b}. The helium is burned to intermediate mass elements. The helium detonation compresses the material as the shock runs over it and heats it up as shown in the right panel of the second row of Fig.~\ref{fig:he_det}. Note that the helium detonation forms at the same place when the additional burning limiter is applied (for details see Appendix~\ref{app}).

The temperatures and densities of the burning Helium layers are not sufficiently hot nor dense enough to also burn the adjacent pure CO material, so the helium detonation first propagates outwards into the helium shell of the primary WD and then starts sweeping around it. At the same time the helium detonation sends a shock-wave into the CO core of the primary WD. A summary of the global properties of the detonation is shown in Table~\ref{tab:det}.

\begin{table*}
\begin{centering}
\begin{tabular}{cccccccccc}
\hline
  Event &  {$\rm T_\mathrm{ign}$} &  {$\rm \rho_\mathrm{ign}$} & {$\rm t_\mathrm{ign}$} & {$\Delta{Q}_{\rm nuc}$} & {$\rm IME$} & {$\rm IGE$}\tabularnewline

- & $[K]$ & $[g\ cm^{-3}]$ & $[sec]$ & $[erg]$ & $[M_\odot]$ & $[M_\odot]$\tabularnewline
\hline 
He det & $2.2\times 10^{9}$ & $2\times 10^{5} $ & $440.5$ & $1.2\times10^{50}$ & $4.7\times10^{-2}$ & $3.5\times10^{-3}$\tabularnewline
C det & $3.8\times 10^{9}$ & $2\times 10^{7}$ & $443.5$ & $4\times10^{50}$ & $3.4\times 10^{-1}$ & $2.3\times10^{-2}$ \tabularnewline
\hline 
\end{tabular}
\par\end{centering}
\caption{Physical conditions that lead to ignition of the helium and carbon detonations and their global yields. The table shows the ignition temperature, ignition density, and ignition time of the detonations as well as the total energy release and the amount of intermediate mass elements and iron group elements that the detonation synthesises.}
\label{tab:det}
\end{table*}

\section{Explosion}
\label{sec:explosion}

After the formation of the helium detonation the situation is initially similar to the well-known double detonation scenario \citep{Fink2007,Guillochon2010,Pakmor2013}, in which the helium detonation travels around the primary WD while sending a shock wave into the carbon-oxygen core. In this scenario the eventually spherical shock-wave converges in a single point in the core where it ignites a carbon detonation that disrupts the WD.

The main difference to the system we investigate here is that the helium detonation does not detonate our primary WD, but detonates the secondary WD instead. Moreover, the secondary WD produces radioactive $^{56}\mathrm{Ni}$ despite its low mass as it is strongly compressed prior to its explosion.

Since the helium detonation will travel all around the primary WD, it is unavoidable that the shock wave that it sends into the CO core will eventually converge at a single point. The strength of the shock, the position of this point, and most importantly the gas density at the convergence point will decide if a carbon detonation forms. The position of the convergence point depends on the speed of the shock wave that is moving through the CO core relative to the speed of the helium detonation moving around the core and the size of the core relative to the circumference of the CO core.

If the helium detonation is much faster than the shock wave it generates at the surface that is traveling into the core, the shock wave starts roughly at the same time everywhere on the surface. In this case the convergence point is close to the center of the core. In contrast, if the helium detonation is comparable in speed or slower than the shock wave that is propagating into the core, the convergence point will be close to the interface between the helium shell and CO core opposite to the ignition point of the helium detonation.

Therefore, the position of the convergence point depends fundamentally on the properties of the helium shell and the CO core of the WD. In particular, a higher density at the base of the helium shell increases the speed and completeness of the nuclear burning and the helium detonation becomes stronger and faster, and also sends a stronger shock into the core. As a rule of thumb, if the shock wave convergences in the CO core at a density $\geq 10^7\,\mathrm{g\,cm^{-3}}$ a carbon detonation will likely form, while carbon burning will likely not start if the density is below $3\times10^6\,\mathrm{g\,cm^{-3}}$ \citep{Seitenzahl2009,Shen2014}.

\begin{figure*}
\includegraphics[width=\textwidth]{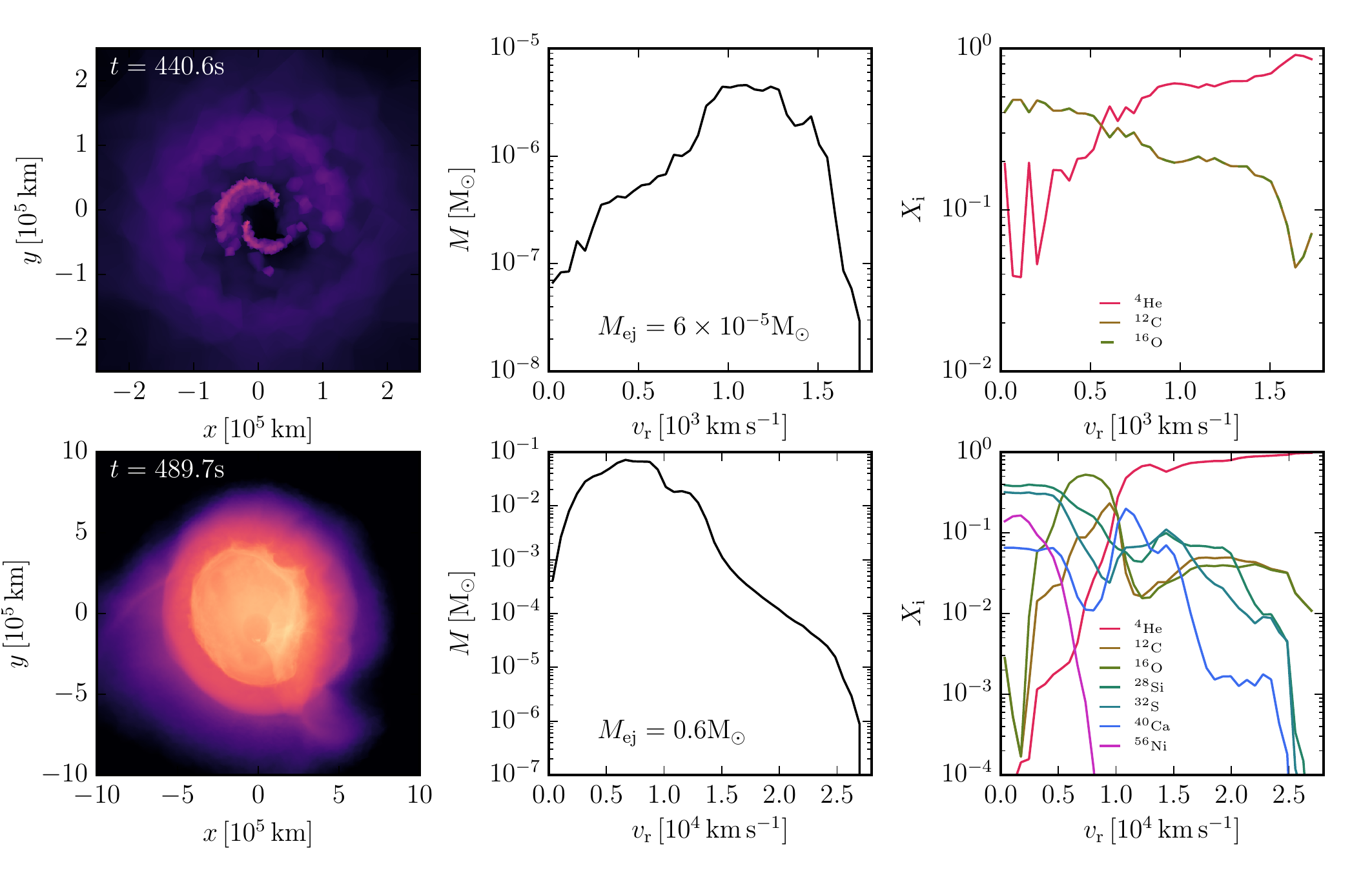}
\caption{Unbound ejecta at the time of the ignition of the helium detonation on the surface of the CO WD (top row) and at the end of the simulation long after the carbon detonation. The left panels show the projected density of the unbound material. The middle and right panels show the distribution of the mass of the ejecta (middle panel) and composition (right panel) in velocity space.}
\label{fig:ejecta}
\end{figure*}



\subsection{Primary WD}

Our primary WD is significantly less massive (only $0.8\,\mathrm{M_\odot}$) than the WDs typically studied in attempts to make type Ia supernovae, which require roughly solar mass primary WDs \citep{Sim2010,Shen2018}. Note also that the central density of our pre-shocked primary WD barely reaches $10^{7}\,\mathrm{g\,cm^{-3}}$, below which a carbon detonation does not produce any radioactive $^{56}\mathrm{Ni}$.

The propagation of the detonation in the helium shell around the primary WD and through its CO core is shown in Fig.~\ref{fig:shock_pri}. It takes the helium detonation about $3\,\mathrm{s}$ to reach the opposite side of the point of ignition. In this time it releases $2\times 10^{49}\,\mathrm{erg}$ from burning the helium shell around the primary WD.

The shock wave it sends into the CO core compresses the core only marginally, temporarily increasing its central density by $\approx 5\%$ from $1.03\times 10^7\,\mathrm{g\,cm^{-3}}$ to $1.08\times 10^7\,\mathrm{g\,cm^{-3}}$.

The shock wave takes about the same time to cross the core as the helium detonation to burn around the core, so that the detonation converges roughly at the edge of the CO core at the opposite side of the ignition of the helium detonation. The shock convergence occurs far off-center at a radius of about $5000\,\mathrm{km}$ and at a density of only $\approx 2\times10^6\,\mathrm{g\ cm^{-3}}$. Owing to the low density and the weak detonation the converging shock fails to ignite a carbon detonation. It is possible that with significantly higher resolution a carbon detonation might form, though the conditions are probably just insufficient as far as we can tell from resolved 1D ignition simulations \citep{Seitenzahl2009,Shen2014}. Note that creating a carbon detonation also fails with an additional burning limiter (for details see Appendix~\ref{app}), so we are reasonably confident that this result does not depend on the details of the numerical treatment of the detonation.

Compared to \citet{Pakmor2013} who simulated a system with a similar primary WD of $1.0\,\mathrm{M_\odot}$ with a helium shell of $0.01\,\mathrm{M_\odot}$ our primary WD in this simulation is less massive, so the central density as well as the density at the base of the helium shell are lower. Therefore the nuclear burning in the helium detonation is less complete, it releases less energy, and the helium detonation is weaker and slower. For comparison, the helium detonation only needs $1\,\mathrm{s}$ to travel around the $1.0\,\mathrm{M_\odot}$ primary WD in \citet{Pakmor2013} compared to $3\,\mathrm{s}$ for our $0.8\,\mathrm{M_\odot}$ primary WD.

One way to improve the chances of creating a carbon detonation may be to increase the strength of the helium detonation, through the existence of a larger He-shell mass. Higher mass could possibly be mediated by a longer period of mass transfer from the secondary hybrid~WD to the primary CO~WD prior to the dynamical interaction between both WDs that we model here. For this the binary system probably needs to transfer several $0.01\,\mathrm{M_\odot}$ of helium to the primary WD during the many orbits in which the secondary WD already fills its Roche lobe but when the accretion rates are very low and the accretion is dynamically unimportant. Whether such a scenario is possible is unknown as, unfortunately, it is not feasible at the moment to properly simulate such a system for the millions of orbits that would be required to model this phase properly. If such a scenario works its rates could be non-negligible \citep{Ruiter2014}. Moreover, the primary CO~WD could also have obtained a He-shell during the evolution of the binary system \citep{Neunteufel2016,Neunteufel2019}.

\subsection{Secondary WD}

The main difference in our system compared to previous simulations \citep[see, e.g.][]{Pakmor2013} is the hybrid nature of the secondary WD, which has a massive helium shell but is more massive than a pure helium WD and has a much higher central density. As we show, this difference gives rise to qualitatively different evolution and outcomes that previous models where He-WD secondaries were considered.  

At the time the helium detonation ignites on the primary, the accretion stream from the secondary WD to the primary WD consists mostly of helium and is degenerate with a density of about $10^5\,\mathrm{g\,cm^{-3}}$. When the helium detonation that is sweeping around the primary WD reaches the end of the accretion stream, it is able to travel upwards through the accretion stream and reach into the helium shell of the hybrid WD. Note that the total mass in the accretion stream is small so its energy release is negligible compared to the helium burned on the surface of the two WDs.

The propagation of the helium detonation around the secondary WD and the shock it sends into its CO core is shown in Fig.~\ref{fig:shock_sec}. Since most of the original $0.1\,\mathrm{M_\odot}$ of helium is still on the secondary WD and the base of the helium shell is at a comparably high density of $\sim 10^6\,\mathrm{g\,cm^{-3}}$ the helium detonation on the secondary WD is very energetic and fast. It sweeps around it in less than $0.5\,\mathrm{s}$, much faster than the shock wave it sends into the CO core of the secondary WD and releases $1.0\times 10^{50}\,\mathrm{erg}$ of energy from burning the helium shell around the secondary WD, about five times the amount released from burning the helium shell around the primary WD.

As shown in Fig.~\ref{fig:shock_sec} the shock wave that is traveling into the core is therefore starting on an almost spherical surface. It converges about $1\,\mathrm{s}$ after the helium detonation around the secondary hybrid WD ceases close to its center. As the shock wave is quite energetic it also compresses the secondary WD in the center prior to converging there and raises its central density significantly by almost a factor of three from $5.6\times10^6\,\mathrm{g\,cm^{-3}}$ to $1.5\times10^7\,\mathrm{g\,cm^{-3}}$.

When the shock wave converges close to the center at a density of about $10^7\,\mathrm{g\,cm^{-3}}$ it heats up carbon enough to ignite a carbon detonation. Similar to the convergence in the primary WD, the formation of the carbon detonation happens independently of the details of the treatment of the detonation (for details see Appendix~\ref{app}). Once the carbon detonation has been ignited, it quickly sweeps through the CO core of the hybrid WD. It releases $4\times 10^{50}\,\mathrm{erg}$ of nuclear energy and unbinds the secondary WD. Its ashes then expand and leave the intact primary WD behind.

\begin{figure*}
\includegraphics[width=\textwidth]{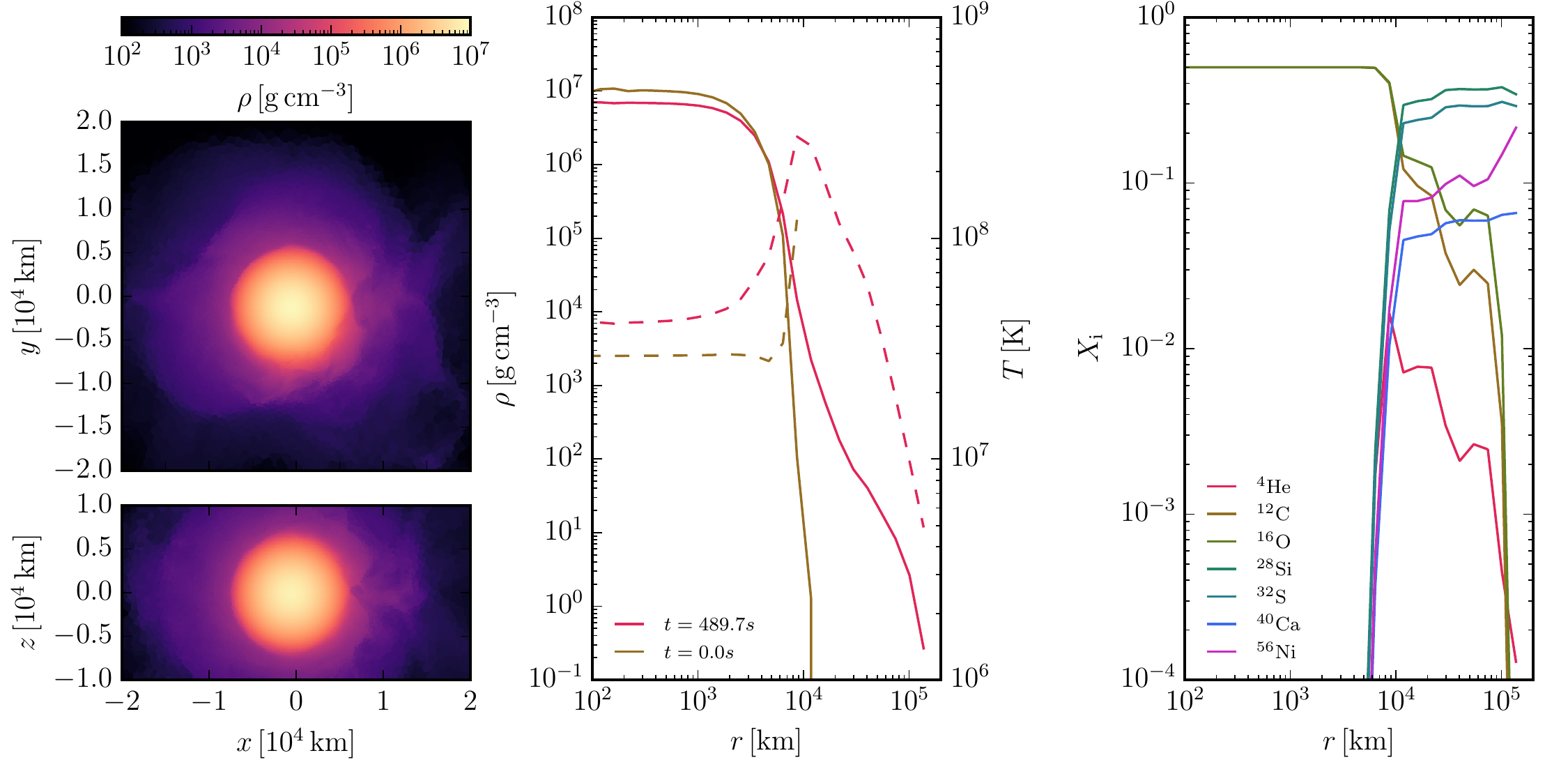}
\caption{Surviving primary WD at the end of the simulation. The left panel shows density slices in the plane of rotation and perpendicular to it. The middle panel shows the radial density (solid lines) and temperature (dashed lines) profile of the surviving WD as well as the initial density profile of the primary WD at the beginning of the simulation for comparison. The right panel shows the composition of the surviving WD.}
\label{fig:remnant}
\end{figure*}

\subsection{The unbound ejecta}

After the carbon detonation has burned the secondary WD completely, its hot ashes become unbound and expand. Eventually they reach homologous expansion. We show the ejecta $46\,\mathrm{s}$ after the ignition of the carbon detonation in the lower row of Fig.~\ref{fig:ejecta}. At this time the structure of the ejecta deviates only by a few percent from homologous expansion. 

The ejecta contain a total of $0.6\,\mathrm{M_\odot}$ which is most of the mass of the secondary WD. The mass distribution of the ejecta is similar to a low energetic type Ia supernova with most of its mass between $5000\,\mathrm{km/s}$ and $10000\,\mathrm{km/s}$ and a very low mass tail up to $27000\,\mathrm{km/s}$. The ejecta consist mostly of oxygen ($0.21\,\mathrm{M_\odot}$), silicon ($0.16\,\mathrm{M_\odot}$), sulfur ($0.09\,\mathrm{M_\odot}$), helium ($0.07\,\mathrm{M_\odot}$), and carbon ($0.06\,\mathrm{M_\odot}$). The ejecta contain only $0.02\,\mathrm{M_\odot}$ of iron group elements of which $0.013\,\mathrm{M_\odot}$ is $^{56}\mathrm{Ni}$. The small amount of iron group elements is a direct consequence of the low central density of the hybrid WD. It produced iron group elements in the carbon detonation only because it was compressed in the center by the helium detonation prior to the ignition of the carbon detonation.

Almost all of the mass of the ejecta is at velocities $v \leq 15000\,\mathrm{km/s}$. The outer parts of the ejecta ($v \geq 10000\,\mathrm{km/s}$) are dominated by unburned helium and only have small trace contributions of carbon, oxygen, and intermediate mass elements. The core of the ejecta ($v < 5000\,\mathrm{km/s}$) is dominated by intermediate mass elements and contains all of the radioactive $^{56}\mathrm{Ni}$ in the ejecta. The part in between, that contains most of the mass is dominated by oxygen and contains significant amounts of intermediate mass elements.

We expect the remnant of this explosion to look similar to the remnant of an ordenary low energy supernova, as it has a similar mass and kinetic energy.

In the top row of Fig.~\ref{fig:ejecta} we show the unbound material at the time the helium detonation ignites, i.e. before any relevant amount of nuclear burning has happened, in a similar way. This material has been ejected from the outer Lagrange point of the secondary WD and forms an outflowing spiral structure around the binary system. It mostly consists of helium and has a typical velocity of $1000\,\mathrm{km/s}$ with a significant tail down to a few $100\,\mathrm{km/s}$ and up to $1500\,\mathrm{km/s}$. Although the ejected mass is unphysically low as we only follow the binary system for a few orbits prior to the explosion the velocity distribution provides an idea of outflowing material in the system prior to the explosion that may become visible either by interaction with the ejecta of the explosion or as early absorption lines in the supernova spectra (see e.g. discussion of CSM interaction in WD-WD mergers in \citealt{Jac+20} and Bobrick et al. in prep.).

We present detailed synthetic light-curves and spectra for the ejecta in Sec.~\ref{sec:observables}.

Note that the ejecta in Fig.~\ref{fig:ejecta} are shown in their rest-frame. This frame moves with a velocity of $v_\mathrm{kick,ejecta}=1600\,\mathrm{km/s}$ relative to the rest-frame of the original binary system. This relative velocity is a direct consequence of burning only one of the WDs. Its material, as it is burned, is moving with the orbital velocity of the secondary WD relative to the center of the binary system. When the ashes of the secondary WD suddenly expand after the nuclear burning has ceased they essentially keep their bulk velocity. Since an external observer sees the binary system and therefore also the ejecta from a random angle, we would expect that spectral lines in observed spectra of this explosion exhibit a shift between $-v_\mathrm{kick,ejecta}$ and $v_\mathrm{kick,ejecta}$ relative to the rest-frame its host galaxy that follows a cosine distribution. The shift is large enough to easily be detected and is a clear prediction for our system.

Note that this velocity shift is inherently expected for any explosion of a single WD in a binary system and its magnitude will be roughly equal to the orbital velocity of the exploding WD. For non-degenerate companions the orbital velocity is typically of order $100\,\mathrm{km/s}$, so this shift will hardly be visible. In contrast, for a binary system of massive CO~WDs \citep{Pakmor2013,Shen2018b} a shift of order of $2000\,\mathrm{km/s}$ is expected that should be detectable in a sufficiently large sample of normal type Ia supernovae if their origin is dominated by such a scenario.
Furthermore, if one identifies a candidate supernova remnant (SNR) from any such scenario in the Galaxy, its center of mass velocity should show such high velocity shifts compared to its rest-frame velocity in the Galaxy. Moreover, if suggested to be related with a hypervelocity WD (e.g. in one of the cases suggested by \citeauthor{Shen2018b} 2018), the SNR center-of-mass velocity should be in the opposite direction from that of the hyper-velocity WD. 


\subsection{The primary WD as surviving hyper-velocity WD}
After the secondary WD suddenly explodes and its ejecta becomes unbound the primary WD is left behind. Without its companion it continues to move on a straight line with the orbital velocity it had at the time of the explosion. For our system it ends up moving with a velocity of $v_\mathrm{kick,WD}=1300\,\mathrm{km/s}$ relative to the rest-frame of the original binary system.

The density profile, temperature profile, and composition profiles of the primary CO~WD at the end of the simulation are shown in Fig.~\ref{fig:remnant}. The core of the WD is essentially undisturbed and very close to spherical for $r \leq 5000\,\mathrm{km}$. The central density has decreased by about a factor of two and the central temperature has increased to about $5\times10^7\,\mathrm{K}$.

At larger radii $r > 5000\,\mathrm{km}$ the WD has changed more significantly. The helium it had accreted prior to the explosion has mostly been burned to heavier elements and became unbound, but some part of it as well as some part of the ejecta of the secondary WD have been captured by the primary WD and now constitute a large fluffy envelope. This envelope has increased the radius of the CO~WD from initially $\sim 10^4\,\mathrm{km}$ by an order of magnitude to $\sim 10^5\,\mathrm{km}$. The envelope is quite hot with a peak temperature of $2\times10^8\,\mathrm{K}$ at the base of the envelope and $10^6\,\mathrm{K}$ at its surface. The composition of the envelope is dominated by the intermediate mass elements with a little bit of helium. There is also a small amount of $5 \times 10^{-3}\,\mathrm{M_\odot}$ radioactive $^{56}\mathrm{Ni}$ present in the envelope.

The surviving WD is slightly more massive than the original primary CO~WD. It lost $0.01\,\mathrm{M_\odot}$ but gained $0.04\,\mathrm{M_\odot}$ from the secondary hybrid WD, most of it from capturing low velocity ejecta of its explosion.

The properties of the surviving WD are particularly interesting in light of the recently found `D6` WDs that have been argued to be the surviving secondary WDs in the dynamically driven double degenerate double detonation (D6) scenario for normal type Ia supernovae \citep{Shen2018b}. These WDs appear to have a large fluffy envelope consisting of intermediate mass and iron group elements. They show neither hydrogen nor helium in their spectra, potentially consistent with our results of no hydrogen and at most very little (not observable) amount of Helium on the surviving WD. In the D6 model, the donor WD must have had a Helium envelope. \citet{Shen2018b} suggest that helium is not observed in the hyper-velocity objects because of low temperatures. It is not clear what the temperatures of the donor should be in the D6 model, but in any case we note that any helium shell might have burned following the explosion, similar to our model where the helium detonation propagates back to the donor. 

The observed WDs move with a velocity of $\sim 10^3\,\mathrm{km}$ relative to the Milky~Way, comparable to our findings. Moreover, their number is significantly lower than what we naively expected to find if every normal type Ia supernova produces one of them, as suggested by the D6 model. These numbers, however, are potentially consistent with the rates we infer. We will discuss this connection further in Sec.~\ref{sec:popsynth} where we attempt to estimate the frequency of events like ours.

\begin{figure}
\includegraphics[width=0.98\linewidth]{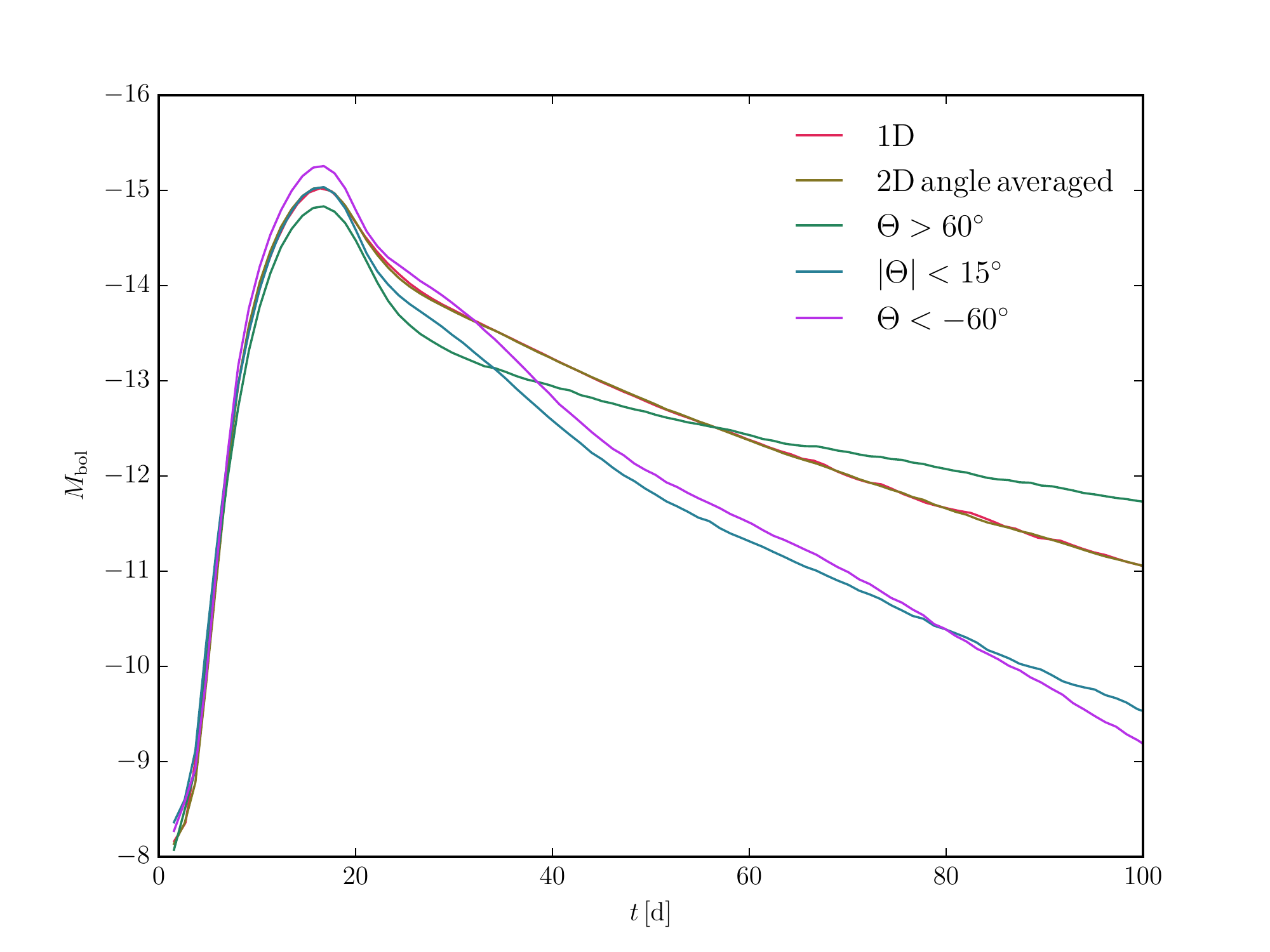}\caption{Bolometric light curves for the spherically symmetric 1D model (red), the angle averaged light curve of the 2D model (brown) and three different angles including both polar directions and the plane of rotation.}
\label{fig:lc_bol}
\end{figure}



\section{Synthetic observables of the ejecta}
\label{sec:observables}

We map the ejecta including their detailed composition from the nucleosynthesis post-processing of $10^6$ Lagrangian tracer particles to a spherical 1D mesh as well as a cylindrical 2D mesh assuming axisymmetry.

We then run the radiation transfer code SuperNu \citep{Wollaeger2013,Wollaeger2014} in order to calculate synthetic lightcurves and spectra for the explosion. SuperNu uses Implicit Monte Carlo (IMC) and Discrete Diffusion Monte Carlo (DDMC) methods to stochastically solve the special-relativistic radiative transport and diffusion equations to first order in $v/c$ in up to three dimensions. The hybrid IMC and DDMC scheme used in SuperNu makes it computationally efficient in regions with high optical depth. This approach allows SuperNu to solve for energy diffusion with very few approximations, which is very relevant for supernova light curves.

The bolometric light-curves for the 1D RT model, the angle averaged light-curve of the 2D RT model, as well as the light-curves for three different lines of sight of the 2D RT model are shown in Fig.~\ref{fig:lc_bol}. The bolometric light-curve peaks $16$d after the explosion at an absolute magnitude of $M_\mathrm{bol}\approx-15$, which makes it significantly fainter than even the faintest SNe~Ia. The faintness is expected for the tiny amount of radioactive material produced.

The bolometric light-curves show significant line of sight dependence, with a spread of about one magnitude at peak between the brightest line of sight (the negative $z$-axis, along the direction of the angular momentum vector of the binary system) and the faintest line of sight (the positive $z$-axis).
Observers in the plain of rotation see a bolometric light-curve that is similar to the angle averaged bolometric light-curve at peak.

At late times the angle averaged bolometric light-curve is dominated by emission in the direction of the positive $z$-axis (which was the faintest direction at peak).

Since the ejecta contain a large amount of $2\times10^{-3}\,\mathrm{M_\odot}$ of $^{44}\mathrm{Ti}$ the transient will likely appear very red.

\section{Event rates and binary evolution}
\label{sec:popsynth}

In this section we estimate the occurrence rate of explosions similar to the one described in detail above based on binary population synthesis modelling. We use the binary evolution code \texttt{SeBa} \citep{Por96,Too12} to simulate the evolution of three million binaries per model starting from the zero-age main-sequence (ZAMS) until the merger of the double WD system. At every time-step, processes such as stellar winds, mass transfer, angular momentum loss, tides, and gravitational radiation are considered with the appropriate prescriptions. 
\texttt{SeBa} is freely available through the Astrophysics MUlti-purpose Software Environment, or AMUSE \citep[][see also \href{http://amusecode.org/}{{\color{blue}amusecode.org}} ]{Por09, Por18}. 

As the main cause for discrepancies between different binary population synthesis codes is found in the choice of input physics and initial conditions \citep{Too14}, we construct two models (model $\alpha\alpha$ and $\gamma\alpha$) that are typically used in double WD modelling with \texttt{SeBa} \citep[see e.g.][]{Too12,Reb19, Zenati2019}. These models differ with respect to the modelling of unstable mass transfer, i.e. common-envelope (CE) evolution \citep{Iva13}. Generally the CE-phase is modelled on the basis of energy conservation \citep{Pac76, Web84}. In this model orbital energy is consumed to unbind the CE with an efficiency $\alpha_{\rm CE}$ (Eq.\,\ref{eq:alpha-ce}). This recipe is used in model $\alpha\alpha$ for every CE-phase. In our alternative model ($\gamma\alpha$) of CE-evolution, we consider a balance of angular momentum with an efficiency parameter $\gamma$ \citep[Eq.\,\ref{eq:gamma-ce}, ][]{Nel00}. 
The $\gamma$-recipe is used unless the binary contains a compact object or the CE is triggered by a tidal instability (rather than dynamically unstable Roche lobe overflow). More details on the models are given in Appendix\,\ref{app-bps}.

Previous work has already shown that hybrid WDs are common \citep{Zenati2019} and frequently merge with other WDs \citep{PeretsZenati2019}. Here we focus on mergers between a massive hybrid WD ($M_{\rm Hybrid} \gtrsim 0.63M_\odot$) and a CO WD ($M_{\rm Hybrid} <M_{\rm CO}\lesssim 0.85M_{\odot}$). On average they make up about several percents of all double WD mergers, giving an integrated rate of several $10^{-5}$ events per solar mass of created stars over a Hubble time. 

The typical evolution towards the merger consists of several phases of interaction. Generally the CO~WD forms first, afterwards the hybrid~WD is formed. This is the case for $76\%$ of the systems in our default models ($M_{\rm Hybrid}>0.63M_{\odot}$, $m_{\rm CO}<0.85M_{\odot}$), and $66-83\%$ with the variations as described below. 

\begin{figure}
    \centering
    \includegraphics[width=0.98\linewidth]{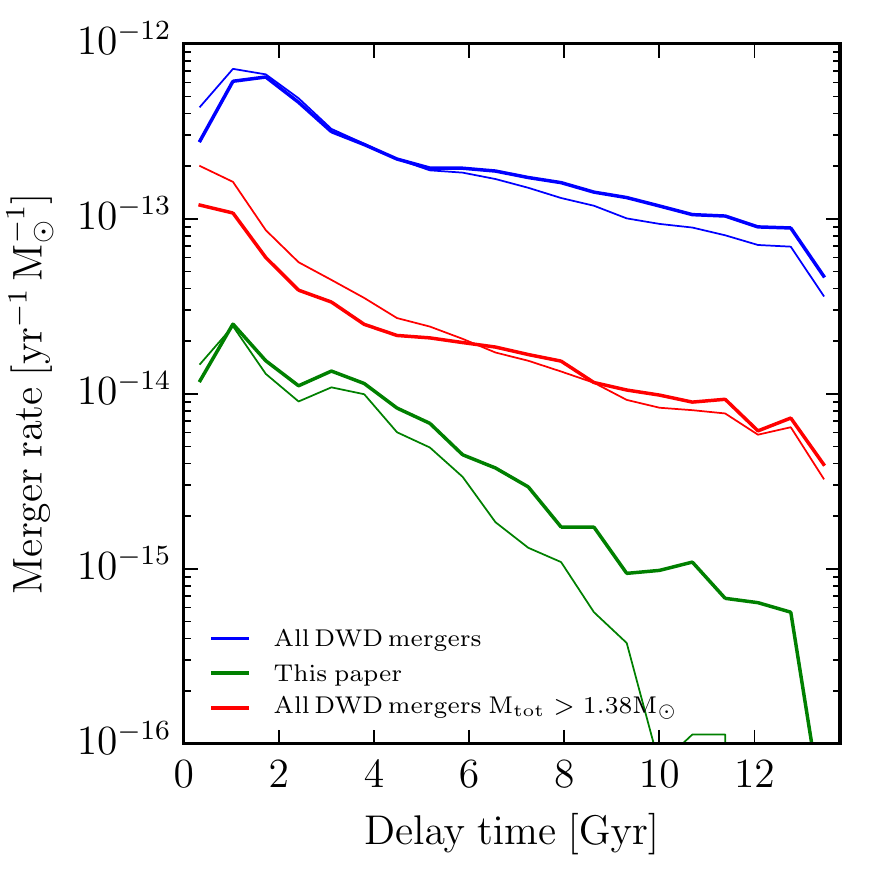}
    \caption{Merger rates for all double WD mergers (DWD) (blue), the traditional DWD merger channel for SNe~Ia that assumes all merging DWD binaries with a total mass $M_\mathrm{tot}>1.38M_\odot$ produce a SN~Ia (red) and the scenario described in this paper (green). Thick lines use the $\gamma\alpha$ model, thin lines the $\alpha\alpha$ model for common envelope evolution.}
    \label{fig:dtd}
\end{figure}

For the majority of these the progenitor of the hybrid is initially (i.e. on the ZAMS) the more massive star in the system. Consequently, it is this star that evolves of the MS before the companion does. It fills its Roche lobe, initiates a phase of stable mass transfer, loses its hydrogen envelope, and becomes a low-mass hydrogen-poor helium-burning star (i.e. stripped star). During the mass transfer the companion has accreted a mass of $\sim 1-2 M_{\odot}$. After the mass transfer phase has ended and the companion has evolved off the MS, the companion initiates a common-envelope phase. After the companion's hydrogen envelope is lost from the system, the binary consists of two stripped stars. At this stage the companion is more massive than the progenitor of the hybrid, due to the previous phase of mass accretion. Consequently its evolutionary timescale as a stripped star is shorter than that of the hybrid progenitor, and it becomes a CO~WD first, then after that the hybrid WD is finally formed. 
This channel is referred to as a formation reversal channel in \citep[][see their section 4.3]{Too12}.

In more detail, in our two models ($\gamma\alpha$ and $\alpha\alpha$) the total merger rate of double WDs is $3.1\times 10^{-3} M_{\odot}^{-1}$  and $3.2\times 10^{-3} M_{\odot}^{-1}$ respectively. Assuming that these events only occur for systems with $M_{\rm Hybrid}>0.63 M_{\odot}$ and $M_{\rm CO}<0.85M_{\odot}$, the synthetic event rate is $ (7-8.5)\times 10^{-5} M_{\odot}^{-1}$.

The event rate is not very sensitive to the exact minimum hybrid mass and maximum CO mass. If the minimum hybrid mass to ensure a scenario as described in this paper is as high as $M_{\rm Hybrid}>0.68$ as is the case for our specific simulation described above, the rate slightly decreases to $(2.9-3.4)\times 10^{-5} M_{\odot}^{-1}$. If instead, the minimum hybrid mass can be as low as $M_{\rm Hybrid}>0.58M_{\odot}$\footnote{We also studied a 3D model for the case of a 0.58 $M_{\odot}$ hybrid, in which case, the hybrid did not detonate, but rather was disrupted later on by the primary, as we shall discuss in more detail in a future paper. It therefore provides us with a lower limit for these companion-detonation SNe}, the rate increases somewhat to $ (1.2-1.5)\times 10^{-4} M_{\odot}^{-1}$. On the other hand, if the mass of the CO~WD can be as high as $0.9M_{\odot}$ the event rates of our default models increases to $ (7.8-9.4)\times 10^{-5} M_{\odot}^{-1}$ (and up to $(1.4-1.6)\times 10^{-4} M_{\odot}^{-1}$ for $M_{\rm Hybrid}>0.58M_{\odot}$).
When reducing the minimum CO mass $M_{\rm CO}<0.8 M_{\odot}$, the event rate decreases slightly to 
$(6.5-7.3)\times 10^{-5} M_{\odot}^{-1}$ (and down to
$(2.7-2.9)\times 10^{-5} M_{\odot}^{-1}$ for $M_{\rm Hybrid}>0.68 M_{\odot}$). 

Overall these types of mergers comprise about 2.5$\%$ of all double WD mergers in our default models, and  0.9-5.1$\%$ with the variations in the limiting values of $M_{\rm Hybrid}$ and $M_{\rm CO}$. 

The event rates described above are based on stellar simulations at Solar metallicity, here taken as $Z=Z_{\odot} = 0.02$.
At lower metallicities the synthetic event rates are not significantly different. The overall rate for the default models $\gamma\alpha$ and $\alpha\alpha$ is $(9.5-9.8)\times 10^{-5} M_{\odot}^{-1}$ at $Z=0.001$.

The event rate is about an order of magnitude lower than the estimated SNe~Ia rate \citep{Li2011, Mao17}. If all SNe~Ia would originate from the D6 scenario, we expect to find $\approx 20$ hypervelocity WDs that are surviving companion WDs of SNe~Ia. However, \citet{Shen2018b} only found three candidates, inconsistent with the D6 scenario but roughly consistent with optimistic rates for our scenario. A significantly larger number of those WDs would likely make them inconsistent with a common origin from our scenario.

\section{Summary and Outlook}
\label{sec:conclusion}

We presented a 3D hydrodynamical simulation of the final phase of a double WD binary consisting of a hybrid~WD with massive $0.1\,\mathrm{M_\odot}$ He shell and a CO~WD that are about to merge.

We find that after some initial mass transfer of helium from the secondary hybrid~WD on the primary CO~WD a thin helium shell builds up around the CO~WD. We showed that as the accretion stream becomes denser and its impact on the surface of the CO~WD becomes more violent eventually a helium detonation forms on the surface of the CO~WD.

The helium detonation wraps around the CO~WD but fails to ignite the CO core as the little amount of helium around the CO~WD only creates a weak detonation and the shock wave from the detonation converges far off-center in the CO~WD at low densities.

However, the helium detonation also travels up the accretion stream and burns the thick helium shell around the hybrid WD. We show that this generates a strong shock wave that converges close to the center of the CO core of the hybrid WD and ignites a carbon detonation.

The hybrid WD is completely burned and unbound by the carbon detonation. Owing to compression of the core of the hybrid WD by the strong shock wave the carbon detonation is able to synthesis $0.018\,\mathrm{M_\odot}$ of $^{56}\mathrm{Ni}$ despite the initial low central density of the hybrid WD. The ejecta lead to a very faint and likely very red transient, that lasts for several tens of days.
We estimated the event rate of the scenario described here as up to $10\%$ of the SNe~Ia rate, making them an interesting target for more sensitive observations by future facilities like the Vera C. Rubin Observatory.

The CO~WD remains intact and is flung at high velocity to become a hyper-velocity WD, with ejection velocity of the order of its original orbital velocity of $1300\mathrm{km/s}$ relative to the rest frame of the original binary system. It collects a thin outer layer from the ashes of the explosion of the hybrid WD that contains $5\times10^{-3}\,\mathrm{M_\odot}$ of $^{56}\mathrm{Ni}$ and provides an alternative origin for identified hyper-velocity WDs to the D6 model proposed by \citep{Shen2018b}. Moreover, the expected rates of such hyper-velocity WDs from our modelled scenario are consistent with the current observationally inferred rate of hyper-velocity-WDs (using the GAIA catalogue; \citet{Shen2018b}), which are significantly lower than suggested by the D6 model.

We expect the center-of-mass velocity of the ejecta of the exploding (secondary) WD to also have a velocity shift of the order of $1600\,\mathrm{km\,s^{-1}}$, in the opposite direction to that of the hyper-velocity WD, which might be observable as a systematic shift of SNe velocities compared with their host galaxies. In Galactic cases, one might find SNRs related to specific hyper-velocity WDs. In those cases the SNR center-of-mass velocity should be similarly high and directed in the opposite direction with that of the hyper-velocity WD.

Explosions of secondary WDs in double WD binary systems have been seen previously \citep{Pakmor2012,Papish2015,Tanikawa2019} though always triggered by an explosion of the primary WD, that is absent in our scenario.

We conclude that detonation of hybrid WDs mediated by He-detonations on primary companions generate new interesting scenarios for novel transients that need to further investigated in 3D hydrodynamical simulations, and searched for by future surveys.

\section*{Data availability}
The simulations underlying this article will be shared on reasonable request to the corresponding author.

\section*{Acknowledgements}

RP thanks Markus Kromer, Stefan Taubenberger, Wolfgang Hillebrandt, and Sasha Kozyreva for interesting and helpful discussions. ST acknowledge support from the Netherlands Research Council NWO (VENI 639.041.645 grants). YZ and HBP thank Daan Van Rossum for interesting discussions and acknowledge support for this project from the European Union's Horizon 2020 research and innovation program under grant agreement No 865932-ERC-SNeX.




\bibliographystyle{mnras}

\begin{thebibliography}{}
\makeatletter
\relax
\def\mn@urlcharsother{\let\do\@makeother \do\$\do\&\do\#\do\^\do\_\do\%\do\~}
\def\mn@doi{\begingroup\mn@urlcharsother \@ifnextchar [ {\mn@doi@}
  {\mn@doi@[]}}
\def\mn@doi@[#1]#2{\def\@tempa{#1}\ifx\@tempa\@empty \href
  {http://dx.doi.org/#2} {doi:#2}\else \href {http://dx.doi.org/#2} {#1}\fi
  \endgroup}
\def\mn@eprint#1#2{\mn@eprint@#1:#2::\@nil}
\def\mn@eprint@arXiv#1{\href {http://arxiv.org/abs/#1} {{\tt arXiv:#1}}}
\def\mn@eprint@dblp#1{\href {http://dblp.uni-trier.de/rec/bibtex/#1.xml}
  {dblp:#1}}
\def\mn@eprint@#1:#2:#3:#4\@nil{\def\@tempa {#1}\def\@tempb {#2}\def\@tempc
  {#3}\ifx \@tempc \@empty \let \@tempc \@tempb \let \@tempb \@tempa \fi \ifx
  \@tempb \@empty \def\@tempb {arXiv}\fi \@ifundefined
  {mn@eprint@\@tempb}{\@tempb:\@tempc}{\expandafter \expandafter \csname
  mn@eprint@\@tempb\endcsname \expandafter{\@tempc}}}

\bibitem[\protect\citeauthoryear{{Beuermann}, {Burwitz}, {Reinsch}, {Schwope}
  \& {Thomas}}{{Beuermann} et~al.}{2020}]{Beuermann20}
{Beuermann} K.,  {Burwitz} V.,  {Reinsch} K.,  {Schwope} A.,   {Thomas} H.~C.,
  2020, \mn@doi [\aap] {10.1051/0004-6361/201936626}, \href
  {https://ui.adsabs.harvard.edu/abs/2020A&A...634A..91B} {634, A91}

\bibitem[\protect\citeauthoryear{{Camacho}, {Torres}, {Garc{\'\i}a-Berro},
  {Zorotovic}, {Schreiber}, {Rebassa-Mansergas}, {Nebot G{\'o}mez-Mor{\'a}n}
  \& {G{\"a}nsicke}}{{Camacho} et~al.}{2014}]{Cam14}
{Camacho} J.,  {Torres} S.,  {Garc{\'\i}a-Berro} E.,  {Zorotovic} M.,
  {Schreiber} M.~R.,  {Rebassa-Mansergas} A.,  {Nebot G{\'o}mez-Mor{\'a}n} A.,
   {G{\"a}nsicke} B.~T.,  2014, \mn@doi [\aap] {10.1051/0004-6361/201323052},
  \href {https://ui.adsabs.harvard.edu/abs/2014A&A...566A..86C} {566, A86}

\bibitem[\protect\citeauthoryear{{Cyburt} et~al.,}{{Cyburt}
  et~al.}{2010}]{Cyburt2010}
{Cyburt} R.~H.,  et~al., 2010, \mn@doi [\apjs] {10.1088/0067-0049/189/1/240},
  \href {http://adsabs.harvard.edu/abs/2010ApJS..189..240C} {189, 240}

\bibitem[\protect\citeauthoryear{{Dewi} \& {Tauris}}{{Dewi} \&
  {Tauris}}{2000}]{Dew00}
{Dewi} J.~D.~M.,  {Tauris} T.~M.,  2000, \aap, \href
  {https://ui.adsabs.harvard.edu/abs/2000A&A...360.1043D} {360, 1043}

\bibitem[\protect\citeauthoryear{{Fink}, {Hillebrandt}  \& {R{\"o}pke}}{{Fink}
  et~al.}{2007}]{Fink2007}
{Fink} M.,  {Hillebrandt} W.,   {R{\"o}pke} F.~K.,  2007, \mn@doi [\aap]
  {10.1051/0004-6361:20078438}, \href
  {https://ui.adsabs.harvard.edu/abs/2007A&A...476.1133F} {476, 1133}

\bibitem[\protect\citeauthoryear{{Guillochon}, {Dan}, {Ramirez-Ruiz}  \&
  {Rosswog}}{{Guillochon} et~al.}{2010}]{Guillochon2010}
{Guillochon} J.,  {Dan} M.,  {Ramirez-Ruiz} E.,   {Rosswog} S.,  2010, \mn@doi
  [\apjl] {10.1088/2041-8205/709/1/L64}, \href
  {https://ui.adsabs.harvard.edu/abs/2010ApJ...709L..64G} {709, L64}

\bibitem[\protect\citeauthoryear{{Iben} \& {Tutukov}}{{Iben} \&
  {Tutukov}}{1985}]{Ibe85}
{Iben} Jr. I.,  {Tutukov} A.~V.,  1985, \mn@doi [\apjs] {10.1086/191054}, \href
  {http://adsabs.harvard.edu/abs/1985ApJS...58..661I} {58, 661}

\bibitem[\protect\citeauthoryear{{Iben}, {Nomoto}, {Tornambe}  \&
  {Tutukov}}{{Iben} et~al.}{1987}]{Iben+87}
{Iben} Icko J.,  {Nomoto} K.,  {Tornambe} A.,   {Tutukov} A.~V.,  1987, \mn@doi
  [\apj] {10.1086/165318}, \href
  {https://ui.adsabs.harvard.edu/abs/1987ApJ...317..717I} {317, 717}

\bibitem[\protect\citeauthoryear{{Istrate}, {Marchant}, {Tauris}, {Langer},
  {Stancliffe}  \& {Grassitelli}}{{Istrate} et~al.}{2016}]{Ist+16}
{Istrate} A.~G.,  {Marchant} P.,  {Tauris} T.~M.,  {Langer} N.,  {Stancliffe}
  R.~J.,   {Grassitelli} L.,  2016, \mn@doi [\aap]
  {10.1051/0004-6361/201628874}, \href
  {http://adsabs.harvard.edu/abs/2016A%26A...595A..35I} {595, A35}

\bibitem[\protect\citeauthoryear{{Ivanova} et~al.,}{{Ivanova}
  et~al.}{2013}]{Iva13}
{Ivanova} N.,  et~al., 2013, \mn@doi [\aapr] {10.1007/s00159-013-0059-2}, \href
  {http://adsabs.harvard.edu/abs/2013A%26ARv..21...59I} {21, 59}

\bibitem[\protect\citeauthoryear{{Jacobson-Gal{\'a}n}
  et~al.,}{{Jacobson-Gal{\'a}n} et~al.}{2020}]{Jac+20}
{Jacobson-Gal{\'a}n} W.~V.,  et~al., 2020, \mn@doi [\apj]
  {10.3847/1538-4357/ab9e66}, \href
  {https://ui.adsabs.harvard.edu/abs/2020ApJ...898..166J} {898, 166}

\bibitem[\protect\citeauthoryear{{Katz} \& {Zingale}}{{Katz} \&
  {Zingale}}{2019}]{Katz2019}
{Katz} M.~P.,  {Zingale} M.,  2019, \mn@doi [\apj] {10.3847/1538-4357/ab0c00},
  \href {https://ui.adsabs.harvard.edu/abs/2019ApJ...874..169K} {874, 169}

\bibitem[\protect\citeauthoryear{{Kupfer} et~al.,}{{Kupfer}
  et~al.}{2020}]{Kupfer20}
{Kupfer} T.,  et~al., 2020, \mn@doi [\apj] {10.3847/1538-4357/ab72ff}, \href
  {https://ui.adsabs.harvard.edu/abs/2020ApJ...891...45K} {891, 45}

\bibitem[\protect\citeauthoryear{{Kushnir} \& {Katz}}{{Kushnir} \&
  {Katz}}{2020}]{Kushnir2020}
{Kushnir} D.,  {Katz} B.,  2020, \mn@doi [\mnras] {10.1093/mnras/staa594},
  \href {https://ui.adsabs.harvard.edu/abs/2020MNRAS.493.5413K} {493, 5413}

\bibitem[\protect\citeauthoryear{{Kushnir}, {Katz}, {Dong}, {Livne}  \&
  {Fern{\'a}ndez}}{{Kushnir} et~al.}{2013}]{Kushnir2013}
{Kushnir} D.,  {Katz} B.,  {Dong} S.,  {Livne} E.,   {Fern{\'a}ndez} R.,  2013,
  \mn@doi [\apjl] {10.1088/2041-8205/778/2/L37}, \href
  {https://ui.adsabs.harvard.edu/abs/2013ApJ...778L..37K} {778, L37}

\bibitem[\protect\citeauthoryear{{Law-Smith} et~al.,}{{Law-Smith}
  et~al.}{2020}]{Law20}
{Law-Smith} J. A.~P.,  et~al., 2020, arXiv e-prints, \href
  {https://ui.adsabs.harvard.edu/abs/2020arXiv201106630L} {p. arXiv:2011.06630}

\bibitem[\protect\citeauthoryear{{Li}, {Chornock}, {Leaman}, {Filippenko},
  {Poznanski}, {Wang}, {Ganeshalingam}  \& {Mannucci}}{{Li}
  et~al.}{2011}]{Li2011}
{Li} W.,  {Chornock} R.,  {Leaman} J.,  {Filippenko} A.~V.,  {Poznanski} D.,
  {Wang} X.,  {Ganeshalingam} M.,   {Mannucci} F.,  2011, \mn@doi [\mnras]
  {10.1111/j.1365-2966.2011.18162.x}, \href
  {https://ui.adsabs.harvard.edu/abs/2011MNRAS.412.1473L} {412, 1473}

\bibitem[\protect\citeauthoryear{{Livio} \& {Soker}}{{Livio} \&
  {Soker}}{1988}]{Liv88}
{Livio} M.,  {Soker} N.,  1988, \mn@doi [\apj] {10.1086/166419}, \href
  {https://ui.adsabs.harvard.edu/abs/1988ApJ...329..764L} {329, 764}

\bibitem[\protect\citeauthoryear{{Loveridge}, {van der Sluys}  \&
  {Kalogera}}{{Loveridge} et~al.}{2011}]{Lov11}
{Loveridge} A.~J.,  {van der Sluys} M.~V.,   {Kalogera} V.,  2011, \mn@doi
  [\apj] {10.1088/0004-637X/743/1/49}, \href
  {https://ui.adsabs.harvard.edu/abs/2011ApJ...743...49L} {743, 49}

\bibitem[\protect\citeauthoryear{{Maoz} \& {Graur}}{{Maoz} \&
  {Graur}}{2017}]{Mao17}
{Maoz} D.,  {Graur} O.,  2017, \mn@doi [\apj] {10.3847/1538-4357/aa8b6e}, \href
  {https://ui.adsabs.harvard.edu/abs/2017ApJ...848...25M} {848, 25}

\bibitem[\protect\citeauthoryear{{Nelemans}, {Verbunt}, {Yungelson}  \&
  {Portegies Zwart}}{{Nelemans} et~al.}{2000}]{Nel00}
{Nelemans} G.,  {Verbunt} F.,  {Yungelson} L.~R.,   {Portegies Zwart} S.~F.,
  2000, \aap, \href {http://adsabs.harvard.edu/abs/2000A%26A...360.1011N} {360,
  1011}

\bibitem[\protect\citeauthoryear{{Nelemans}, {Yungelson}, {Portegies Zwart}  \&
  {Verbunt}}{{Nelemans} et~al.}{2001}]{Nel01}
{Nelemans} G.,  {Yungelson} L.~R.,  {Portegies Zwart} S.~F.,   {Verbunt} F.,
  2001, \mn@doi [\aap] {10.1051/0004-6361:20000147}, \href
  {https://ui.adsabs.harvard.edu/abs/2001A&A...365..491N} {365, 491}

\bibitem[\protect\citeauthoryear{{Neunteufel}, {Yoon}  \&
  {Langer}}{{Neunteufel} et~al.}{2016}]{Neunteufel2016}
{Neunteufel} P.,  {Yoon} S.~C.,   {Langer} N.,  2016, \mn@doi [\aap]
  {10.1051/0004-6361/201527845}, \href
  {https://ui.adsabs.harvard.edu/abs/2016A&A...589A..43N} {589, A43}

\bibitem[\protect\citeauthoryear{{Neunteufel}, {Yoon}  \&
  {Langer}}{{Neunteufel} et~al.}{2019}]{Neunteufel2019}
{Neunteufel} P.,  {Yoon} S.~C.,   {Langer} N.,  2019, \mn@doi [\aap]
  {10.1051/0004-6361/201935322}, \href
  {https://ui.adsabs.harvard.edu/abs/2019A&A...627A..14N} {627, A14}

\bibitem[\protect\citeauthoryear{{Ohlmann}, {R{\"o}pke}, {Pakmor}  \&
  {Springel}}{{Ohlmann} et~al.}{2017}]{Ohlmann2017}
{Ohlmann} S.~T.,  {R{\"o}pke} F.~K.,  {Pakmor} R.,   {Springel} V.,  2017,
  \mn@doi [\aap] {10.1051/0004-6361/201629692}, \href
  {https://ui.adsabs.harvard.edu/abs/2017A&A...599A...5O} {599, A5}

\bibitem[\protect\citeauthoryear{{Paczynski}}{{Paczynski}}{1976}]{Pac76}
{Paczynski} B.,  1976, in {P.~Eggleton, S.~Mitton, \& J.~Whelan} ed.,  IAU
  Symposium Vol. 73, Structure and Evolution of Close Binary Systems. pp 75--+

\bibitem[\protect\citeauthoryear{{Pakmor}, {Bauer}  \& {Springel}}{{Pakmor}
  et~al.}{2011}]{Pakmor2011b}
{Pakmor} R.,  {Bauer} A.,   {Springel} V.,  2011, \mn@doi [\mnras]
  {10.1111/j.1365-2966.2011.19591.x}, \href
  {http://adsabs.harvard.edu/abs/2011MNRAS.418.1392P} {418, 1392}

\bibitem[\protect\citeauthoryear{{Pakmor}, {Edelmann}, {R{\"o}pke}  \&
  {Hillebrand t}}{{Pakmor} et~al.}{2012}]{Pakmor2012}
{Pakmor} R.,  {Edelmann} P.,  {R{\"o}pke} F.~K.,   {Hillebrand t} W.,  2012,
  \mn@doi [\mnras] {10.1111/j.1365-2966.2012.21383.x}, \href
  {https://ui.adsabs.harvard.edu/abs/2012MNRAS.424.2222P} {424, 2222}

\bibitem[\protect\citeauthoryear{{Pakmor}, {Kromer}, {Taubenberger}  \&
  {Springel}}{{Pakmor} et~al.}{2013}]{Pakmor2013}
{Pakmor} R.,  {Kromer} M.,  {Taubenberger} S.,   {Springel} V.,  2013, \mn@doi
  [\apjl] {10.1088/2041-8205/770/1/L8}, \href
  {http://adsabs.harvard.edu/abs/2013ApJ...770L...8P} {770, L8}

\bibitem[\protect\citeauthoryear{{Pakmor}, {Springel}, {Bauer}, {Mocz},
  {Munoz}, {Ohlmann}, {Schaal}  \& {Zhu}}{{Pakmor} et~al.}{2016}]{Pakmor2016}
{Pakmor} R.,  {Springel} V.,  {Bauer} A.,  {Mocz} P.,  {Munoz} D.~J.,
  {Ohlmann} S.~T.,  {Schaal} K.,   {Zhu} C.,  2016, \mn@doi [\mnras]
  {10.1093/mnras/stv2380}, \href
  {http://adsabs.harvard.edu/abs/2016MNRAS.455.1134P} {455, 1134}

\bibitem[\protect\citeauthoryear{{Papish}, {Soker}, {Garc{\'\i}a-Berro}  \&
  {Aznar-Sigu{\'a}n}}{{Papish} et~al.}{2015}]{Papish2015}
{Papish} O.,  {Soker} N.,  {Garc{\'\i}a-Berro} E.,   {Aznar-Sigu{\'a}n} G.,
  2015, \mn@doi [\mnras] {10.1093/mnras/stv337}, \href
  {https://ui.adsabs.harvard.edu/abs/2015MNRAS.449..942P} {449, 942}

\bibitem[\protect\citeauthoryear{{Parsons} et~al.,}{{Parsons}
  et~al.}{2020}]{Steven20}
{Parsons} S.~G.,  et~al., 2020, \mn@doi [Nature Astronomy]
  {10.1038/s41550-020-1037-z}, \href
  {https://ui.adsabs.harvard.edu/abs/2020NatAs.tmp...56P} {}

\bibitem[\protect\citeauthoryear{{Paxton}, {Bildsten}, {Dotter}, {Herwig},
  {Lesaffre}  \& {Timmes}}{{Paxton} et~al.}{2011}]{Paxton2011}
{Paxton} B.,  {Bildsten} L.,  {Dotter} A.,  {Herwig} F.,  {Lesaffre} P.,
  {Timmes} F.,  2011, \mn@doi [\apjs] {10.1088/0067-0049/192/1/3}, \href
  {https://ui.adsabs.harvard.edu/abs/2011ApJS..192....3P} {192, 3}

\bibitem[\protect\citeauthoryear{{Paxton} et~al.,}{{Paxton}
  et~al.}{2019}]{Paxton2019}
{Paxton} B.,  et~al., 2019, \mn@doi [\apjs] {10.3847/1538-4365/ab2241}, \href
  {https://ui.adsabs.harvard.edu/abs/2019ApJS..243...10P} {243, 10}

\bibitem[\protect\citeauthoryear{{Perets}, {Zenati}, {Toonen}  \&
  {Bobrick}}{{Perets} et~al.}{2019}]{PeretsZenati2019}
{Perets} H.~B.,  {Zenati} Y.,  {Toonen} S.,   {Bobrick} A.,  2019, arXiv
  e-prints, \href {https://ui.adsabs.harvard.edu/abs/2019arXiv191007532P} {p.
  arXiv:1910.07532}

\bibitem[\protect\citeauthoryear{{Portegies Zwart}}{{Portegies
  Zwart}}{2013}]{Por13}
{Portegies Zwart} S.,  2013, \mn@doi [\mnras] {10.1093/mnrasl/sls022}, \href
  {https://ui.adsabs.harvard.edu/abs/2013MNRAS.429L..45P} {429, L45}

\bibitem[\protect\citeauthoryear{{Portegies Zwart} \& {McMillan}}{{Portegies
  Zwart} \& {McMillan}}{2018}]{Por18}
{Portegies Zwart} S.,  {McMillan} S.,  2018, {Astrophysical Recipes; The art of
  AMUSE}, \mn@doi{10.1088/978-0-7503-1320-9.
}

\bibitem[\protect\citeauthoryear{{Portegies Zwart} \& {Verbunt}}{{Portegies
  Zwart} \& {Verbunt}}{1996}]{Por96}
{Portegies Zwart} S.~F.,  {Verbunt} F.,  1996, \aap, \href
  {http://adsabs.harvard.edu/abs/1996A\%26A...309..179P} {309, 179}

\bibitem[\protect\citeauthoryear{{Portegies Zwart} et~al.,}{{Portegies Zwart}
  et~al.}{2009}]{Por09}
{Portegies Zwart} S.,  et~al., 2009, \mn@doi [\na]
  {10.1016/j.newast.2008.10.006}, \href
  {http://adsabs.harvard.edu/abs/2009NewA...14..369P} {14, 369}

\bibitem[\protect\citeauthoryear{{Prada Moroni} \& {Straniero}}{{Prada Moroni}
  \& {Straniero}}{2009}]{Moroni09}
{Prada Moroni} P.~G.,  {Straniero} O.,  2009, \mn@doi [\aap]
  {10.1051/0004-6361/200912847}, \href
  {https://ui.adsabs.harvard.edu/abs/2009A&A...507.1575P} {507, 1575}

\bibitem[\protect\citeauthoryear{{Rebassa-Mansergas}, {Toonen}, {Korol}  \&
  {Torres}}{{Rebassa-Mansergas} et~al.}{2019}]{Reb19}
{Rebassa-Mansergas} A.,  {Toonen} S.,  {Korol} V.,   {Torres} S.,  2019,
  \mn@doi [\mnras] {10.1093/mnras/sty2965}, \href
  {https://ui.adsabs.harvard.edu/abs/2019MNRAS.482.3656R} {482, 3656}

\bibitem[\protect\citeauthoryear{{Ruiter}, {Belczynski}, {Sim}, {Seitenzahl}
  \& {Kwiatkowski}}{{Ruiter} et~al.}{2014}]{Ruiter2014}
{Ruiter} A.~J.,  {Belczynski} K.,  {Sim} S.~A.,  {Seitenzahl} I.~R.,
  {Kwiatkowski} D.,  2014, \mn@doi [\mnras] {10.1093/mnrasl/slu030}, \href
  {https://ui.adsabs.harvard.edu/abs/2014MNRAS.440L.101R} {440, L101}

\bibitem[\protect\citeauthoryear{{Sato}, {Nakasato}, {Tanikawa}, {Nomoto},
  {Maeda}  \& {Hachisu}}{{Sato} et~al.}{2015}]{Sato+15}
{Sato} Y.,  {Nakasato} N.,  {Tanikawa} A.,  {Nomoto} K.,  {Maeda} K.,
  {Hachisu} I.,  2015, \mn@doi [\apj] {10.1088/0004-637X/807/1/105}, \href
  {https://ui.adsabs.harvard.edu/abs/2015ApJ...807..105S} {807, 105}

\bibitem[\protect\citeauthoryear{{Seitenzahl}, {Meakin}, {Townsley}, {Lamb}  \&
  {Truran}}{{Seitenzahl} et~al.}{2009}]{Seitenzahl2009}
{Seitenzahl} I.~R.,  {Meakin} C.~A.,  {Townsley} D.~M.,  {Lamb} D.~Q.,
  {Truran} J.~W.,  2009, \mn@doi [\apj] {10.1088/0004-637X/696/1/515}, \href
  {https://ui.adsabs.harvard.edu/abs/2009ApJ...696..515S} {696, 515}

\bibitem[\protect\citeauthoryear{{Shen} \& {Bildsten}}{{Shen} \&
  {Bildsten}}{2014}]{Shen2014}
{Shen} K.~J.,  {Bildsten} L.,  2014, \mn@doi [\apj]
  {10.1088/0004-637X/785/1/61}, \href
  {https://ui.adsabs.harvard.edu/abs/2014ApJ...785...61S} {785, 61}

\bibitem[\protect\citeauthoryear{{Shen} \& {Moore}}{{Shen} \&
  {Moore}}{2014}]{Shen2014b}
{Shen} K.~J.,  {Moore} K.,  2014, \mn@doi [\apj] {10.1088/0004-637X/797/1/46},
  \href {https://ui.adsabs.harvard.edu/abs/2014ApJ...797...46S} {797, 46}

\bibitem[\protect\citeauthoryear{{Shen}, {Kasen}, {Miles}  \&
  {Townsley}}{{Shen} et~al.}{2018a}]{Shen2018}
{Shen} K.~J.,  {Kasen} D.,  {Miles} B.~J.,   {Townsley} D.~M.,  2018a, \mn@doi
  [\apj] {10.3847/1538-4357/aaa8de}, \href
  {https://ui.adsabs.harvard.edu/abs/2018ApJ...854...52S} {854, 52}

\bibitem[\protect\citeauthoryear{{Shen} et~al.,}{{Shen}
  et~al.}{2018b}]{Shen2018b}
{Shen} K.~J.,  et~al., 2018b, \mn@doi [\apj] {10.3847/1538-4357/aad55b}, \href
  {https://ui.adsabs.harvard.edu/abs/2018ApJ...865...15S} {865, 15}

\bibitem[\protect\citeauthoryear{{Sim}, {R{\"o}pke}, {Hillebrandt}, {Kromer},
  {Pakmor}, {Fink}, {Ruiter}  \& {Seitenzahl}}{{Sim} et~al.}{2010}]{Sim2010}
{Sim} S.~A.,  {R{\"o}pke} F.~K.,  {Hillebrandt} W.,  {Kromer} M.,  {Pakmor} R.,
   {Fink} M.,  {Ruiter} A.~J.,   {Seitenzahl} I.~R.,  2010, \mn@doi [\apjl]
  {10.1088/2041-8205/714/1/L52}, \href
  {https://ui.adsabs.harvard.edu/abs/2010ApJ...714L..52S} {714, L52}

\bibitem[\protect\citeauthoryear{{Springel}}{{Springel}}{2010}]{Arepo}
{Springel} V.,  2010, \mn@doi [\mnras] {10.1111/j.1365-2966.2009.15715.x},
  \href {http://adsabs.harvard.edu/abs/2010MNRAS.401..791S} {401, 791}

\bibitem[\protect\citeauthoryear{{Tanikawa}, {Nomoto}, {Nakasato}  \&
  {Maeda}}{{Tanikawa} et~al.}{2019}]{Tanikawa2019}
{Tanikawa} A.,  {Nomoto} K.,  {Nakasato} N.,   {Maeda} K.,  2019, \mn@doi
  [\apj] {10.3847/1538-4357/ab46b6}, \href
  {https://ui.adsabs.harvard.edu/abs/2019ApJ...885..103T} {885, 103}

\bibitem[\protect\citeauthoryear{{Timmes} \& {Swesty}}{{Timmes} \&
  {Swesty}}{2000}]{Timmes2000}
{Timmes} F.~X.,  {Swesty} F.~D.,  Timmes2000, \mn@doi [\apjs] {10.1086/313304},
  \href {http://adsabs.harvard.edu/abs/2000ApJS..126..501T} {126, 501}

\bibitem[\protect\citeauthoryear{{Toonen} \& {Nelemans}}{{Toonen} \&
  {Nelemans}}{2013}]{Too13}
{Toonen} S.,  {Nelemans} G.,  2013, \mn@doi [\aap]
  {10.1051/0004-6361/201321753}, \href
  {https://ui.adsabs.harvard.edu/abs/2013A&A...557A..87T} {557, A87}

\bibitem[\protect\citeauthoryear{{Toonen}, {Nelemans}  \& {Portegies
  Zwart}}{{Toonen} et~al.}{2012}]{Too12}
{Toonen} S.,  {Nelemans} G.,   {Portegies Zwart} S.,  2012, \mn@doi [\aap]
  {10.1051/0004-6361/201218966}, \href
  {http://adsabs.harvard.edu/abs/2012A%26A...546A..70T} {546, A70}

\bibitem[\protect\citeauthoryear{{Toonen}, {Claeys}, {Mennekens}  \&
  {Ruiter}}{{Toonen} et~al.}{2014}]{Too14}
{Toonen} S.,  {Claeys} J.~S.~W.,  {Mennekens} N.,   {Ruiter} A.~J.,  2014,
  \mn@doi [\aap] {10.1051/0004-6361/201321576}, \href
  {http://adsabs.harvard.edu/abs/2014A%26A...562A..14T} {562, A14}

\bibitem[\protect\citeauthoryear{{Toonen}, {Hollands}, {G{\"a}nsicke}  \&
  {Boekholt}}{{Toonen} et~al.}{2017}]{Too17}
{Toonen} S.,  {Hollands} M.,  {G{\"a}nsicke} B.~T.,   {Boekholt} T.,  2017,
  \mn@doi [\aap] {10.1051/0004-6361/201629978}, \href
  {https://ui.adsabs.harvard.edu/abs/2017A&A...602A..16T} {602, A16}

\bibitem[\protect\citeauthoryear{{Tutukov} \& {Yungelson}}{{Tutukov} \&
  {Yungelson}}{1992}]{Tut+92}
{Tutukov} A.~V.,  {Yungelson} L.~R.,  1992, \sovast, \href
  {http://adsabs.harvard.edu/abs/1992SvA....36..266T} {36, 266}

\bibitem[\protect\citeauthoryear{{Webbink}}{{Webbink}}{1984}]{Web84}
{Webbink} R.~F.,  1984, \mn@doi [\apj] {10.1086/161701}, \href
  {http://adsabs.harvard.edu/abs/1984ApJ...277..355W} {277, 355}

\bibitem[\protect\citeauthoryear{{Weinberger}, {Springel}  \&
  {Pakmor}}{{Weinberger} et~al.}{2019}]{Weinberger2020}
{Weinberger} R.,  {Springel} V.,   {Pakmor} R.,  2019, arXiv e-prints, \href
  {https://ui.adsabs.harvard.edu/abs/2019arXiv190904667W} {p. arXiv:1909.04667}

\bibitem[\protect\citeauthoryear{{Wollaeger} \& {van Rossum}}{{Wollaeger} \&
  {van Rossum}}{2014}]{Wollaeger2014}
{Wollaeger} R.~T.,  {van Rossum} D.~R.,  2014, \mn@doi [\apjs]
  {10.1088/0067-0049/214/2/28}, \href
  {https://ui.adsabs.harvard.edu/abs/2014ApJS..214...28W} {214, 28}

\bibitem[\protect\citeauthoryear{{Wollaeger}, {van Rossum}, {Graziani},
  {Couch}, {Jordan}, {Lamb}  \& {Moses}}{{Wollaeger}
  et~al.}{2013}]{Wollaeger2013}
{Wollaeger} R.~T.,  {van Rossum} D.~R.,  {Graziani} C.,  {Couch} S.~M.,
  {Jordan} IV G.~C.,  {Lamb} D.~Q.,   {Moses} G.~A.,  2013, \mn@doi [\apjs]
  {10.1088/0067-0049/209/2/36}, \href
  {https://ui.adsabs.harvard.edu/abs/2013ApJS..209...36W} {209, 36}

\bibitem[\protect\citeauthoryear{{Xu} \& {Li}}{{Xu} \& {Li}}{2010}]{Xu10}
{Xu} X.-J.,  {Li} X.-D.,  2010, \mn@doi [\apj] {10.1088/0004-637X/716/1/114},
  \href {https://ui.adsabs.harvard.edu/abs/2010ApJ...716..114X} {716, 114}

\bibitem[\protect\citeauthoryear{{Zenati}, {Toonen}  \& {Perets}}{{Zenati}
  et~al.}{2019}]{Zenati2019}
{Zenati} Y.,  {Toonen} S.,   {Perets} H.~B.,  2019, \mn@doi [\mnras]
  {10.1093/mnras/sty2723}, \href
  {https://ui.adsabs.harvard.edu/abs/2019MNRAS.482.1135Z} {482, 1135}

\bibitem[\protect\citeauthoryear{{Zhang}, {Hall}, {Jeffery}  \& {Bi}}{{Zhang}
  et~al.}{2018}]{Zha+18}
{Zhang} X.,  {Hall} P.~D.,  {Jeffery} C.~S.,   {Bi} S.,  2018, \mn@doi [\mnras]
  {10.1093/mnras/stx2747}, \href
  {http://adsabs.harvard.edu/abs/2018MNRAS.474..427Z} {474, 427}

\bibitem[\protect\citeauthoryear{{Zorotovic}, {Schreiber}, {G{\"a}nsicke}  \&
  {Nebot G{\'o}mez-Mor{\'a}n}}{{Zorotovic} et~al.}{2010}]{Zor10}
{Zorotovic} M.,  {Schreiber} M.~R.,  {G{\"a}nsicke} B.~T.,   {Nebot
  G{\'o}mez-Mor{\'a}n} A.,  2010, \mn@doi [\aap] {10.1051/0004-6361/200913658},
  \href {https://ui.adsabs.harvard.edu/abs/2010A&A...520A..86Z} {520, A86}

\bibitem[\protect\citeauthoryear{{de Kool}, {van den Heuvel}  \& {Pylyser}}{{de
  Kool} et~al.}{1987}]{deK87}
{de Kool} M.,  {van den Heuvel} E.~P.~J.,   {Pylyser} E.,  1987, \aap, \href
  {https://ui.adsabs.harvard.edu/abs/1987A&A...183...47D} {183, 47}

\makeatother
\end{thebibliography}


\appendix
\section{Nuclear burning limiter}
\label{app}

\begin{figure}
\includegraphics[width=0.65\linewidth]{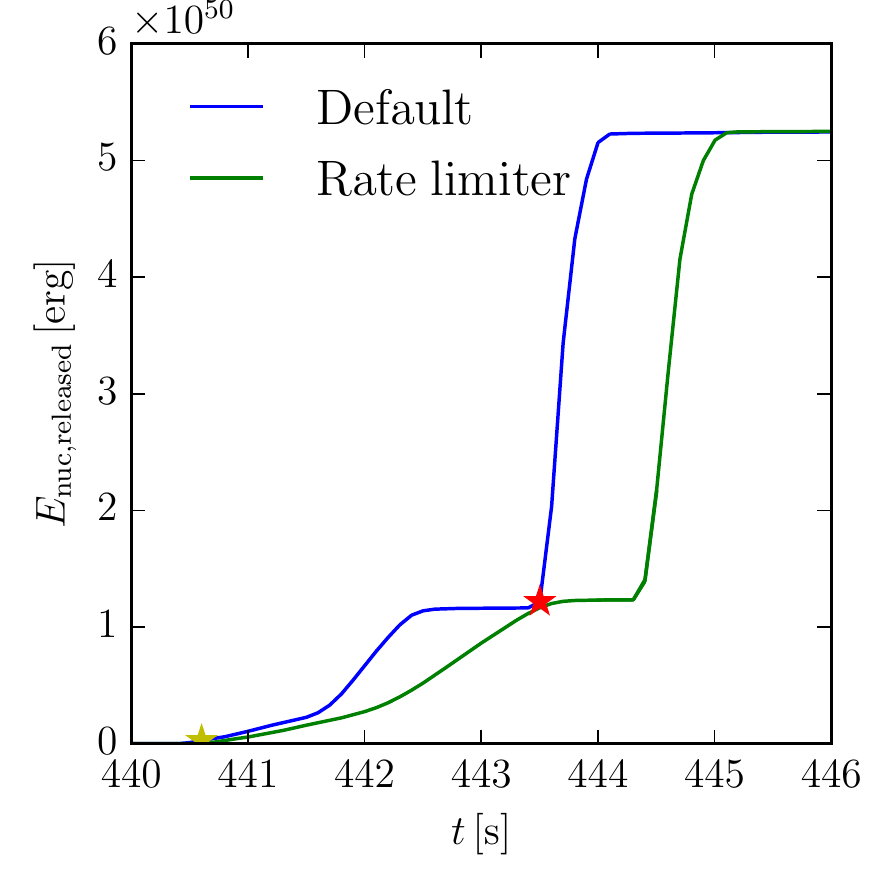}
\caption{Nuclear energy release with time for the default simulation (blue) discussed in the main paper and an otherwise identical simulation using an additional limiter for the nuclear burning that reduces nuclear reaction rates (green). The yellow and red star denote the times when the helium and carbon detonations ignite for the default simulation.}
\label{fig:enuc_limiter}
\end{figure}

\begin{figure*}
\includegraphics[width=0.8\textwidth]{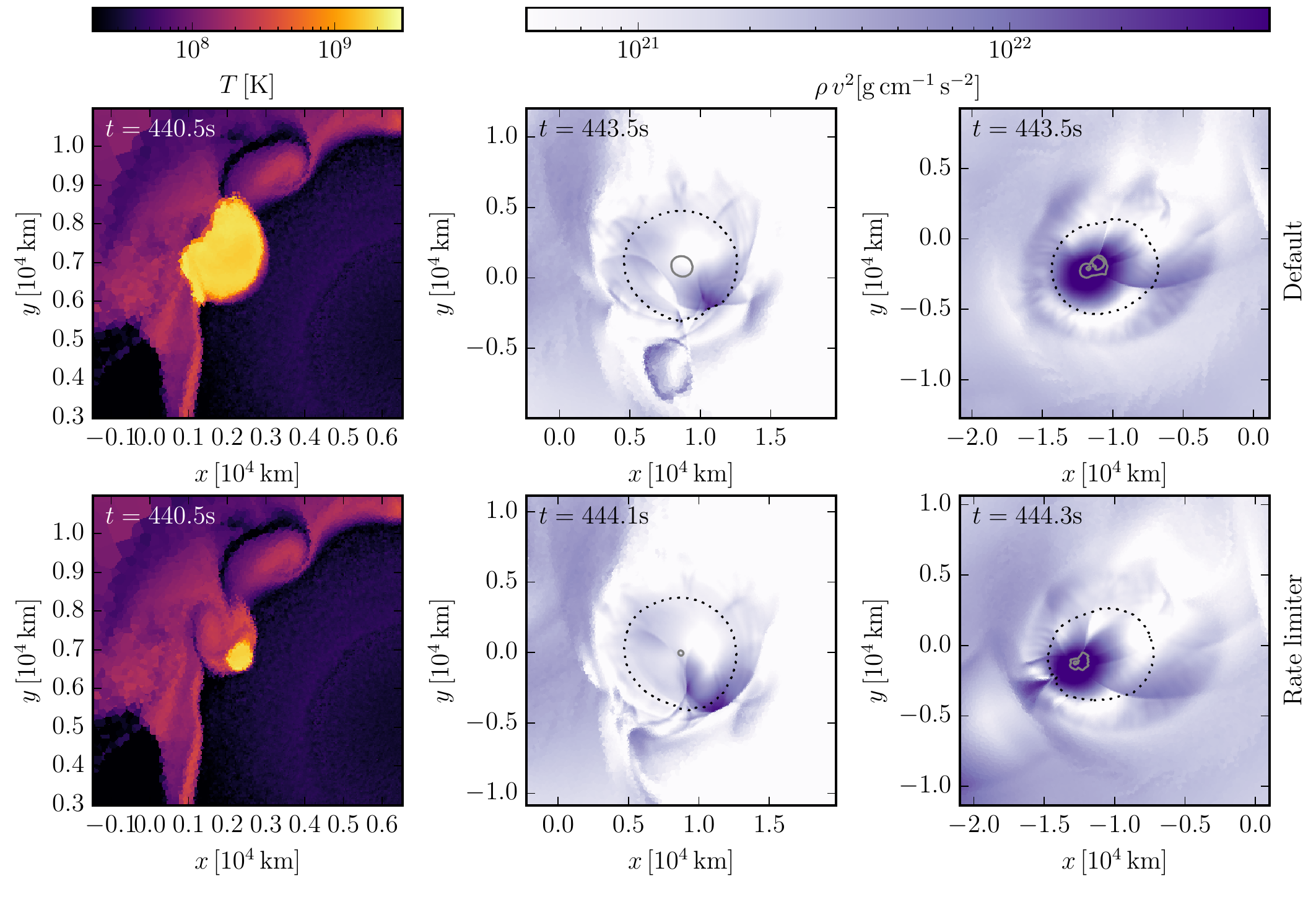}
\caption{Comparison of the ignition of the helium detonation and the convergence of the shock fronts in both WDs of the default simulation (top row) with an otherwise identical simulation using an additional limiter for the nuclear burning that reduces nuclear reaction rates (bottom row). The left panel shows the temperature around the ignition point of the helium detonation $0.1$s after ignition. The middle and right panels show the primary and secondary WD, respectively, at the times when the shockwaves that originate from the helium detonation converge in their CO core.}
\label{fig:limiter}
\end{figure*}

Recently it has been discussed that the nuclear burning limiter that is needed in 3D simulations of thermonuclear explosions because detonations are fundamentally unresolved can make a difference for yields of the nuclear burning \citep{Kushnir2013,Shen2018,Katz2019,Kushnir2020}.

To better the understand the differences between limiters as well as improve our confidence in our results we have repeated the merger simulation in \textsc{arepo} until nuclear burning ceases with a limiter on the nuclear reaction rates \citep{Kushnir2013}.

As described in the main text, in our default configuration we attempt to detect the cells directly behind the shock of the detonation and disable nuclear burning in them. In practice we disable nuclear burning for cells that fullfill $\nabla \cdot \vec{v} < 0$ and $\left| \nabla P \right| r_\mathrm{cell} / P > 0.66$.

In our rerun we dynamically reduce all nuclear reaction rates in a cell by a constant factor such that $\Delta \ln T \leq 0.1$ in every timestep. This ensures that the nuclear timescale is always longer than the hydrodynamical timescale of a cell.

We show the time evolution of the nuclear energy release of the standard simulation (the same as in the lower panel Fig.~\ref{fig:evolution}) and the rerun with the additional reaction rate limiter in Fig.~\ref{fig:enuc_limiter}. Both simulations show qualitatively the same evolution.

As shown in Fig.~\ref{fig:limiter} they both ignite a helium detonation at the same time and place on the surface of the CO~WD. Moreover, in both simulations the helium detonation ignites a carbon detonation only in the hybrid WD. The total nuclear energy release of the helium detonation as well as the carbon detonation is essentially identical. Thus we conclude that our main results are independent of the details of the nuclear burning limiter.

The main difference is the speed of the energy release during the helium detonation. In the simulation with the additional rate limiter the nuclear reactions are significantly slowed down, so that the helium detonation releases its energy more slowly. This also slows down the propagation of the helium detonation and the carbon detonation is ignited about $1\,\mathrm{s}$ later compared to our standard simulation without the rate limiter. The difference is much smaller for the carbon detonation, though it is also slowed down a little bit by the rate limiter. Nevertheless, the final yields and total energy release are the same for either limiter making us confident that our results are at least qualitatively stable.

\section{Common-envelope phase}
\label{app-bps}

In this study we simulate the effect of the common-envelope phase in two ways (Sec.\,\ref{sec:popsynth}); based on energy conservation \citep{Pac76, Web84} or a balance of angular momentum \citep{Nel00}.
In the former model orbital energy is consumed to unbind the CE with an efficiency $\alpha_{\rm CE}$:
\begin{equation}
E_{\rm bin} = \frac{GM_dM_c}{\lambda R} = \alpha_{\rm CE} E_{\rm orbit},
\label{eq:alpha-ce}
\end{equation}
where $E_{\rm bin}$ is the binding energy of the envelope, $M_d$ is the mass of the donor star, $M_c$ the mass of its core, $R$ its radius, and $E_{\rm orbit}$ the orbital energy of the binary at the onset of the common-envelope phase. $\lambda$ the structure parameter of its envelope \citep{Liv88, deK87, Dew00, Xu10, Lov11}. In the model $\alpha\alpha$, we assume $\alpha_{\rm CE}\lambda=2$.
This is calibrated to the formation of double WDs \citep[i.e. second phase of mass transfer, see][]{Nel00, Nel01, Too12}.

In the latter model, the $\gamma$-parameter describes the efficiency with which orbital angular momentum is used to expel the CE according to
\begin{equation}
\frac{J_{\rm b, init}-J_{\rm b,  final}}{J_{\rm b,init}} = \gamma \frac{\Delta M_{\rm d}}{M_{\rm d}+ M_{\rm a}},
\label{eq:gamma-ce}
\end{equation} 
where $J_{\rm b,init}$ and $J_{\rm b,final}$ are the orbital angular momentum of the pre- and post-mass transfer binary respectively, and $M_{\rm a}$ is the mass of the companion. The motivation for the $\gamma$-formalism comes from observed double WD systems. The first phase of mass transfer in the evolution to form a double WD could not be explained by the $\alpha$-formalism nor stable mass transfer for a Hertzsprung gap donor star \citep[see][]{Nel00}.
Here, we assume $\gamma = 1.75$, which is calibrated to the mass-ratio distribution of observed double WDs \citep{Nel01, Too12}.

Other constraints on common-envelope evolution exist which are based on other types of binaries. For example, from the demographics of the observed population of compact WD-M-dwarf systems (often call. ed post-common envelope binaries) a small value of $\alpha_{\rm CE}\lambda(\approx 0.25)$ is derived \citep{Zor10, Too13, Por13, Cam14}. Recently the hydro-dynamical simulation of \citep{Law20} achieved a successful CE ejection and a measurable CE efficiency of 
$\alpha_{\rm CE} \approx 0.1-0.4$ for a CE initiated by a 12M$_{\odot}$ supergiant with a neutron-star companion. However, in the context of double WDs, modelling each CE-phase with such a small value of $\alpha_{\rm CE}\lambda$, leads to a mass ratio distribution that is at odds with the observed population, see e.g \cite{Nel00} and  \cite[][their model $\alpha\alpha2$]{Too17}.

Additionally, we note that the synthetic double WD space density from both our models $\alpha\alpha$ and $\gamma\alpha$ is in agreement with the observed range \citep[][see their discussion, section 6.2]{Reb19}. Despite the uncertainties in binary population synthesis modelling in general, and the common-envelope phase in particular, the match in space densities provides confidence in the synthetic double WD merger rates provided in this paper, especially for model $\gamma\alpha$ that also reproduces the observed mass ratio distribution of double white dwarfs.

\bsp	
\label{lastpage}
\end{document}